\documentclass[onecolumn]{aastex631}

\usepackage{comment}
\usepackage{amsmath}

\pdfoutput=1

\newcommand{\dgg}{^{\circ}}
\newcommand{\dg}{$^{\circ}$ }

\received{May 3, 2024}
\revised{June 7, 2024}
\accepted{July 15, 2024}

\shorttitle{Spacecraft Trajectory Effects on in situ Signatures using 3DCOREweb}
\shortauthors{H.T. Rüdisser et al.}
\graphicspath{{./}{figures/}}

\begin{document}

\title{Understanding the effects of spacecraft trajectories through solar coronal mass ejection flux ropes using 3DCOREweb}

\author[0000-0002-2559-2669]{Hannah T. Rüdisser}
\affiliation{Austrian Space Weather Office, GeoSphere Austria, Graz, Austria}
\affiliation{Institute of Physics, University of Graz, Graz, Austria}
\correspondingauthor{Hannah T. Rüdisser}
\email{hannah.ruedisser@geosphere.at}

\author[0000-0002-6273-4320]{Andreas J. Weiss}
\affiliation{NASA Postdoctoral Program Fellow, NASA Goddard Space Flight Center, Greenbelt, MD, USA}

\author[0000-0001-5387-9512]{Justin Le Louëdec}
\affiliation{Austrian Space Weather Office, GeoSphere Austria, Graz, Austria}

\author[0000-0003-1516-5441]{Ute V. Amerstorfer}
\affiliation{Austrian Space Weather Office, GeoSphere Austria, Graz, Austria}

\author[0000-0001-6868-4152]{Christian Möstl}
\affiliation{Austrian Space Weather Office, GeoSphere Austria, Graz, Austria}

\author[0000-0001-9992-8471]{Emma E. Davies}
\affiliation{Austrian Space Weather Office, GeoSphere Austria, Graz, Austria}

\author[0000-0002-0547-7671]{Helmut Lammer}
\affiliation{Space Research Institute, Austrian Academy of Sciences, Graz, Austria}

\begin{abstract}

This study investigates the impact of spacecraft positioning and trajectory on in situ signatures of coronal mass ejections (CMEs). Employing the 3DCORE model, a 3D flux rope model that can generate in situ profiles for any given point in space and time, we conduct forward modeling to analyze such signatures for various latitudinal and longitudinal positions, with respect to the flux rope apex, at 0.8~au. Using this approach, we explore the appearance of the resulting in situ profiles for different flux rope types, with different handedness and inclination angles, for both high and low twist CMEs. Our findings reveal that CMEs exhibit distinct differences in signatures between apex hits and flank encounters, with the latter displaying elongated profiles with reduced rotation. Notably, constant, non-rotating in situ signatures are only observed for flank encounters of low twist CMEs, suggesting the existence of untwisted magnetic field lines within CME legs. Additionally, our study confirms the unambiguous appearance of different flux rope types in in situ signatures in most of the cases, barring some indistinguishable cases, contributing to the broader understanding and interpretation of observational data. Given the model assumptions, this may refute trajectory effects to be the cause for mismatching flux rope types as identified in solar signatures. While acknowledging limitations inherent in our model, such as the assumption of constant twist and non-deformable torus-like shape, we still draw relevant conclusions within the context of global magnetic field structures of CMEs and the potential for distinguishing flux rope types based on in situ observations.

\end{abstract}


\section{Introduction} \label{sec:intro}

A major unsolved problem in space weather forecasting is the correct prediction of the time series of the solar wind magnetic field vector in near-Earth space. In particular, spacecraft have observed structures in situ in the solar wind inside interplanetary coronal mass ejections (ICMEs), which lead to the strongest excursions in the total magnetic field. ICMEs typically comprise a shock, sheath and magnetic ejecta, where the field vector smoothly rotates up to 180\dg \citep{zurbuchen2006situ}. This is usually interpreted as the observational signature of a spacecraft passing through a large scale magnetic flux rope within the magnetic ejecta \citep{burlaga1981magnetic, marubashi1986structure,burlaga1988magnetic}, in which helical magnetic field lines are wound around a central axis. This is described by the \cite{lundquist1950} or \cite{gold1960origin} solutions of the basic equations of magnetohydrodynamics (MHD). This picture has been generally accepted as the main concept capable of reconstructing the global structure of CMEs, and towards predicting their in situ magnetic field structure for space weather forecasting. Many different models have been developed which are based on the flux rope hypothesis \citep[e.g.][]{lepping1990magnetic,hidalgo2012,isavnin_2016,Moestl_2018,rouillard_2020,nieves_chinchilla_2020}. However, other interpretations have been suggested. One such interpretation is spheromaks, known as possible MHD solutions \citep{Farrugia_1995,vandas2017magnetic}. Another interpretation involves a different kind of rotating magnetic field that gives similarly rotating signatures emerging from simulations of erupting flux ropes \citep{al2013magnetic,al2019evolution}.

\begin{figure*}[h!]
\centering
{\includegraphics[width=\textwidth]{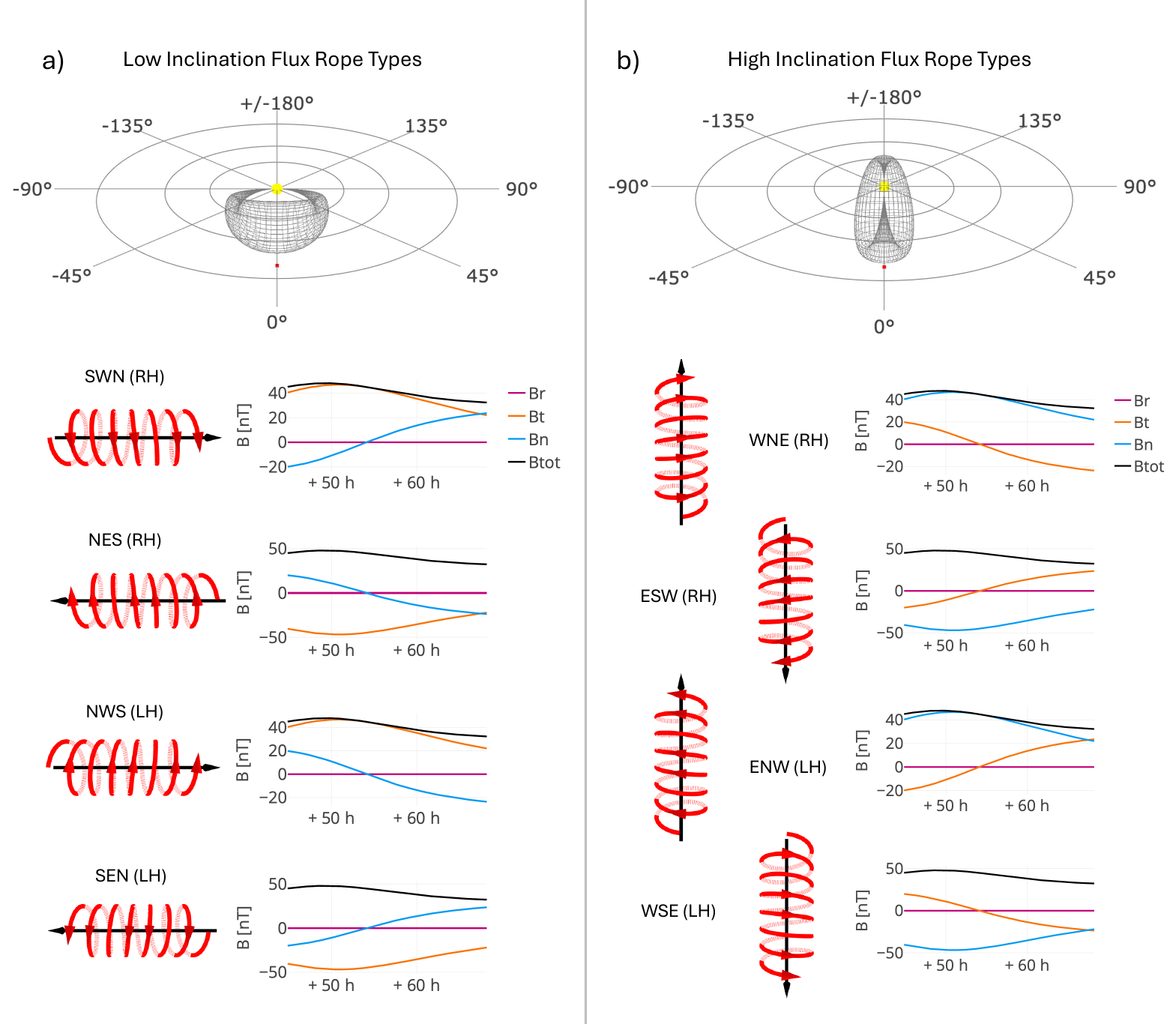}}
\caption{Graphic representation of the eight flux rope types: a) low inclination flux rope types, and b) high inclination flux rope types. A 3DCORE model is shown at the top for both low and high inclinations, and the simplified sketches of the flux ropes together with the in situ profiles for an apex hit for each flux rope type are visualised below. Each flux rope type is labelled next to the corresponding sketch, along with its handedness (LH: left-handed, RH: right-handed).} 
\label{fig:lowinctypes}
\end{figure*}

Following \cite{bothmer1998structure} and \cite{mulligan1998solar}, there exists a comprehensive classification scheme for flux ropes, categorizing them into eight distinct types based on their chirality and orientation in space. This arises from a flux rope being characterized by both the poloidal field wrapping around, and the axial field running parallel to the central axis. Low inclination flux ropes have their central axis nearly parallel to the ecliptic plane, which leads to a change of sign in the normal magnetic field component $B_{n}$ upon crossing. Depending on their handedness and orientation of the central axis, they can be classified as one of the following: left-handed: North-West-South (NWS), South-East-North (SEN); right-handed: North-East-South (NES), South-West-North (SWN). Flux ropes with their axis nearly perpendicular to the ecliptic plane experience a sign change in the tangential component $B_t$ and are termed high inclination flux ropes. The four different types are: left-handed: East-North-West (ENW), West-South-East (WSE); right-handed: East-South-West (ESW), West-North-East (WNE). A graphical representation of the flux rope types is shown in Figure \ref{fig:lowinctypes}.

The global structure of CMEs is believed to have both ends of the flux rope connected to the Sun for a considerable duration during its propagation through the heliosphere. This is suggested by the presence of bidirectional electron flows in some CMEs up to 1~au \citep[e.g.][]{richardson1997,shodhan2000}. Nevertheless, while \cite{zurbuchen2006situ} proposed the legs of a CME to contain highly twisted field lines, \cite{owens2016legs} suggested the existence of ``legs'' of untwisted magnetic flux. This theory might serve as an additional explanation for the appearance of ICME signatures that deviate from the ideal flux rope signatures. 

Although relatively rare, measurements of the same ICME by spacecraft at multiple points over varying heliocentric distances and longitudinal separations have provided valuable insights into their structural characteristics. While there is a general consistency with the helical flux rope (FR) model, it is noteworthy that not all ICMEs exhibit clear FR structures. Different studies have highlighted significant variations among individual ICMEs, deviating from overall statistical trends \citep[for overviews see][]{kilpua2009,good2018correlation,lugaz2018, salman2020radial, davies2022multi}. This variability suggests possible deviations from the classic flux rope hypothesis. These deviations may result from interactions with other structures or the background solar wind, and from different spacecraft measurement points and cuts through the ICME \citep[e.g.][]{Moestl_2009,mostl2012multi}. Such factors particularly affect the flattening of the circular cross section \citep[e.g.][]{riley2004kinematic, owens2006kinematically, savani2010observational, davies2021}. For an overview on these various processes CMEs undergo as they propagate away from the Sun, we refer the reader to \cite{manchester2017physical}, \cite{luhmann2020} and \cite{temmer2023}.

\cite{Regnault_2023} recently highlighted the scarcity of simultaneous multipoint measurements of the same ICME, emphasizing the need for more data to enhance our understanding. Looking ahead, currently operating missions such as Parker Solar Probe, Solar Orbiter, BepiColombo, Wind, and STEREO-A are anticipated to contribute significantly to this field by providing additional multipoint observations \cite[e.g.][]{moestl_2022,lugaz2024}. These missions, currently operational within 1~au, offer opportunities to study a broader range of ICME events and further refine our understanding of their complex structures and dynamics.

A subset of magnetic flux ropes that can neither be fitted using the force-free model \citep{lepping1990magnetic}, nor be reconstructed with the Grad-Shafranov technique \citep{HuSonnerup_2002}, are known as ``magnetic-cloud-like'' \citep[MCL,][]{lepping2006} ejecta. One special observational signature within these that has puzzled researchers for a long time are so-called ``back regions'' of ICMEs. These ``back regions'' follow after the main magnetic field rotation of the flux rope has passed the observer. In these regions, the field stays somewhat elevated, but does rotate only very slightly and the field components remain essentially constant. Two explanations have been put forward for this behaviour: (1) \cite{ruffenach2012multispacecraft} show how magnetic reconnection at the front of a CME is thought to alter the magnetic field signature in the back of the flux rope. Due to the magnetic field lines of the flux rope reconnecting with the interplanetary magnetic field at the front, the in situ signature seems to have a tail or back region with little to no magnetic field rotation but more or less constant components. \cite{dasso2006, dasso2007} show examples of ICME events exhibiting such tails or back regions interpreted as results of magnetic reconnection as the CME propagates through the solar wind. (2) Another explanation is based on the idea that the observer passes through the flux rope in such a way that the constant field arises for a quasi-stationary observer purely from a geometrical effect \citep{moestl_2010} and is actually part of the flux rope itself. In reality, both processes may contribute to the ICME signature that is then observed by a spacecraft. 

Improving modeling capabilities to account for the high complexity of CME propagation and evolution has been a key element of studies within the past years  \citep[e.g.][]{pomoell2018, sarkar2020observationally, pal2022cmemodel, kay2023advancescmemodelling, maharana2023}. An investigation of the influence of initial parameters of a CME MHD simulation on the results at different latitudinal and longitudinal locations at 1~au has been conducted by \cite{shen2021numerical2}. Their study indicated, that as long as the initial mass of the CME remains unchanged, the initial geometric thickness will have a different influence in the latitudinal and longitudinal direction. \cite{Scolin2021} used a swarm of simulated spacecraft in radial alignment to study the development of CME magnetic complexity during propagation. They were able to relate the probability of detecting complexity changes to the interaction with large-scale solar wind structures. Similarly, a swarm of synthetic spacecraft at different latitudinal and longitudinal positions was used in \cite{Scolini_2023} to study the coherence of CMEs. Their simulations suggest a CME to retain a lower complexity and higher coherence along its magnetic axis, while still requiring crossings along both the axial and perpendicular directions for a characterization of its global complexity. Additionally, coherence is found to be lower in the west flank of the structure, due to Parker spiral effects.

A study comparing stationary spacecraft and trajectories similar to Parker Solar Probe in an MHD simulation was conducted by \cite{lynch2022}. Finding in situ flux rope models to be generally applicable for the inference of some physical properties, such as size and flux content, they conclude shortcomings in the determination of the flux rope axis orientation.

In this study, we test two hypotheses: (1) Constant, non-rotating (CNR) in situ signatures can be explained by an effect of the trajectory of an observer through a 3D expanding magnetic flux rope, without invoking magnetic reconnection as an explanation for those signatures. (2) Trajectory effects can lead to a misinterpretation of flux rope types. According to \cite{palmerio2018coronal}, only about half of the studied events show a match regarding their flux rope type, when comparing in situ observations to observations from the Sun. Therefore, we investigate the appearance of different flux rope types for different trajectories and attempt to validate or refute the second hypothesis.

To this end, we employ the 3DCORE model, which is a 3D magnetic field model for ICME flux ropes first presented by \cite{Moestl_2018} and significantly improved by \cite{weiss2021analysis,weiss2021triple}. We place up to 20 synthetic spacecraft at 0.8~au and analyze how their respective positions influence measured in situ signatures. Thereby, we attempt to gain insight into how CNRs present in ICME in situ signatures can be generated.

In Section \ref{sec:method}, we give a short summary of the 3DCORE model in general and report on a new tool called 3DCOREweb, which facilitates its open-source usability. We additionally outline the general setup of the simulation. In Section \ref{sec:results}, we show the in situ signatures obtained by spacecraft at different locations, and discuss these in Section \ref{sec:discussion}.

\section{Method} \label{sec:method}

\subsection{3DCOREweb}

3DCORE is a semi-empirical 3D flux rope model, which is described in detail in \cite{weiss2021analysis,weiss2021triple}. In the past, the model has proven successful in inferring flux rope parameters by fitting in situ magnetic field observations employing an Approximate Bayesian Computation-Sequential Monte Carlo (ABC-SMC) method \citep[e.g.][]{davies2021, telloni2021, moestl_2022, long2023}. In the case of limited longitudinal separations (within $\sim 5 \dgg$), even multipoint events have successfully been fitted simultaneously \citep[e.g.][]{davies2021, weiss2021triple}. 3DCORE assumes a torus-like shape with an embedded Gold-Hoyle-like magnetic field, based on an analytical solution by \cite{vandas2017magnetic}. The structure is attached to the Sun on both ends and expands self-similarly during its propagation. The ambient solar wind is approximated using a constant solar wind speed as one of the models input parameters. The interaction with the background solar wind is considered through a simple drag model \citep{vrsnak2013propagation}, allowing for the frontal part of the CME to either be accelerated or decelerated, depending on the relative speed of the ambient solar wind. By including an elliptical cross-section through variation of the aspect ratio, propagation effects like flattening of the flux rope in the direction of propagation \citep[also known as ``pancaking'',][]{riley2004kinematic} are approximated. A more thorough explanation of the mathematical setup of the model and its 13 input parameters can be found in \cite{weiss2021analysis, weiss2021triple}. These 13 parameters are: longitude, latitude, inclination, diameter at 1~au ($D_{1\mathrm{au}}$), aspect ratio ($\delta$), launch radius ($R_{0}$), launch velocity ($V_{0}$), expansion rate ($n_{a}$), background drag ($\Gamma$), solar wind speed ($V_{sw}$), twist factor ($T_{f}$), magnetic decay rate ($n_b$) and magnetic field strength at 1~au ($B_{1\mathrm{au}}$).  The magnetic field within the toroidal structure is given for custom curvilinear coordinates ($r \in \lceil 0, \infty )$, $\psi \in \lceil 0, 2 \pi )$, $\phi \in \lceil 0, 2 \pi )$) based on the toroidal coordinate system by

\begin{subequations}\label{eq:scalinglaws}
  \begin{align}
    B_r &= 0\\
    B_{\psi} &= \frac{B_0}{1+b^2r^2}\\
    B_{\phi} &= \frac{B_0br}{(1+b^2r^2)\left( 1 + r \frac{\rho_1}{\rho_0} \cos \phi \right)},
  \end{align}
\end{subequations}

following a Gold-Hoyle-like solution described in \cite{vandas2017magnetic}, with $B_0$ as the magnetic field at the central axis of the toroidal structure and $b$ as a modified twist parameter defined as

\begin{equation}\label{eq:twist}
    b = T_{f} \frac{\rho_1}{2\pi\rho_0} \sin \left(\frac{\psi}{2} \right)^2.
\end{equation}

where $\rho_0$ and $\rho_1$ define the major and minor radius of the base torus, while the flux rope structure itself is defined by the implicit volume $r \leq 1$. As defined in \cite{weiss2021triple}, the twist factor $T_{f}$ used as input parameter relates to the number of twists $\tau$ as $\tau = T_f/E(\delta)$, where $E(\delta)$ is a function for the circumference of an ellipse with a given aspect ratio $\delta$. The scaling laws are implemented according to \citet{leitner2007consequences} as

\begin{subequations}\label{eq:magfield}
  \begin{align}
    D(t) &= D_{1au}R_{apex}(t)^{n_a}\\
    B_0(t) &= B_{1au}R_{apex}(t)^{- n_b},
  \end{align}
\end{subequations}

including the time dependency. $R_{apex}$ hereby refers to the distance of the CME's apex point to the Sun. 

While the model's simplicity allows for real-time application as well as the generation of large ensembles, there are some drawbacks which have to be taken into account. Examples of such drawbacks are the fixed flux rope width of $180\dgg$, as well as the background solar wind speed which is assumed to be constant throughout the heliosphere. While this implementation still produces sufficiently reasonable results locally, the application to multipoint events has shown that the global large-scale structure is not adequately and realistically enough modelled to overcome large spacecraft separations \citep[][(in revision, ApJ)]{weiss2021triple, davies2024september}. In the past, this has been circumvented by fitting both in situ spacecraft observations separately and retrospectively comparing their results.  This approach has yielded meaningful conclusions, but still has to be considered when analysing and interpreting model results. Additionally, deformations and deflections aside from the flattening of the cross section as well as a non constant twist have not yet been implemented. Nevertheless, these possible improvements are already subject to further investigation \citep{weiss2022}.

\begin{figure*}[h!]
\centering
{\includegraphics[width=\textwidth]{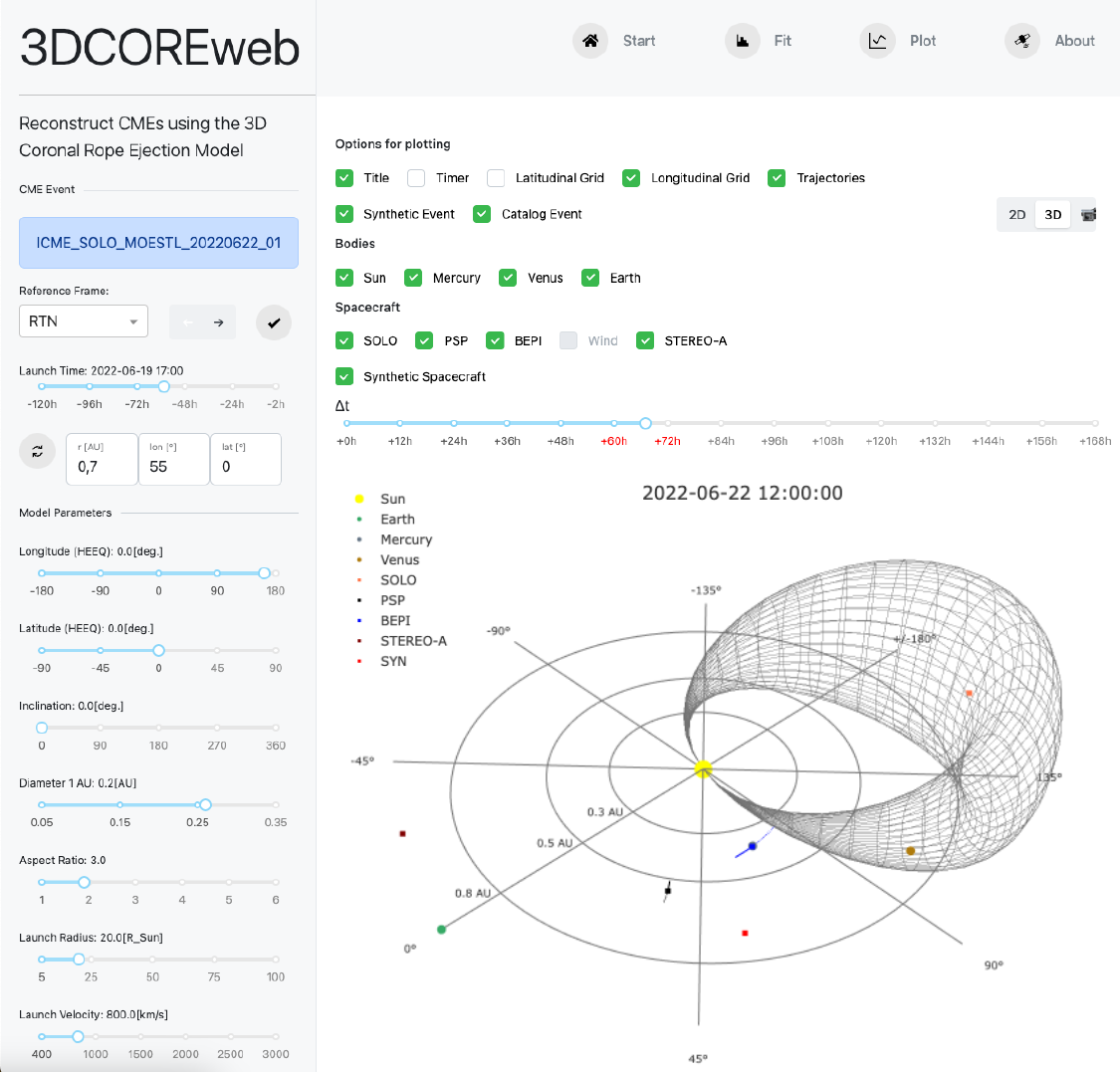}}
\caption{A screenshot of the graphical user interface in 3DCOREweb, as available via GitHub. This screenshot shows the subpage ``Plot'', where the user is able to create interactive 3D plots inserting the 3DCORE model into the heliosphere. The panel on the left shows the details of the event, in this case an ICME measured by Solar Orbiter (SOLO) on 2022-06-22. It also contains the sliders to adjust the model input parameters. Above the plot, the user can modify plotting options and visualize the temporal evolution by adjusting the time slider. Furthermore, the user can visualize generated in situ signatures of the model, use the ABC-SMC algorithm as introduced in \cite{weiss2021analysis}, and analyse the results.} 
\label{fig:webinterface}
\end{figure*}

Recently, we have made an effort to increase the model's usability through 3DCOREweb\footnote{https://github.com/hruedisser/3DCOREweb}. A Dash \citep{plotly} application is available via GitHub and includes a graphical user interface that can be used without any prior Python knowledge. An example screenshot of this graphical user interface is shown in Figure \ref{fig:webinterface}. 

The user can choose to reconstruct ICME signatures from the HELIO4CAST ICME catalog\footnote{https://helioforecast.space/icmecat}, an extensive living catalog of interplanetary coronal mass ejections \citep{moestl2020ICMECAT}. Alternatively, they can select a custom event for various spacecraft (e.g. Solar Orbiter, Parker Solar Probe, STEREO-A) after the data is pre-downloaded from a given repository. The straightforward implementation of the fitting algorithm as well as a clear overview of the results facilitate the application of the tool for the general scientific community. Additionally, we exploited the usage of 3DCORE as a simple forward modeling tool, including a large variety of Python routines for modeling, analyzing and plotting.

\subsection{Simulation Setup} \label{sec:experimentsetup}

\begin{figure*}[h!]
\centering
{\includegraphics[width=\textwidth]{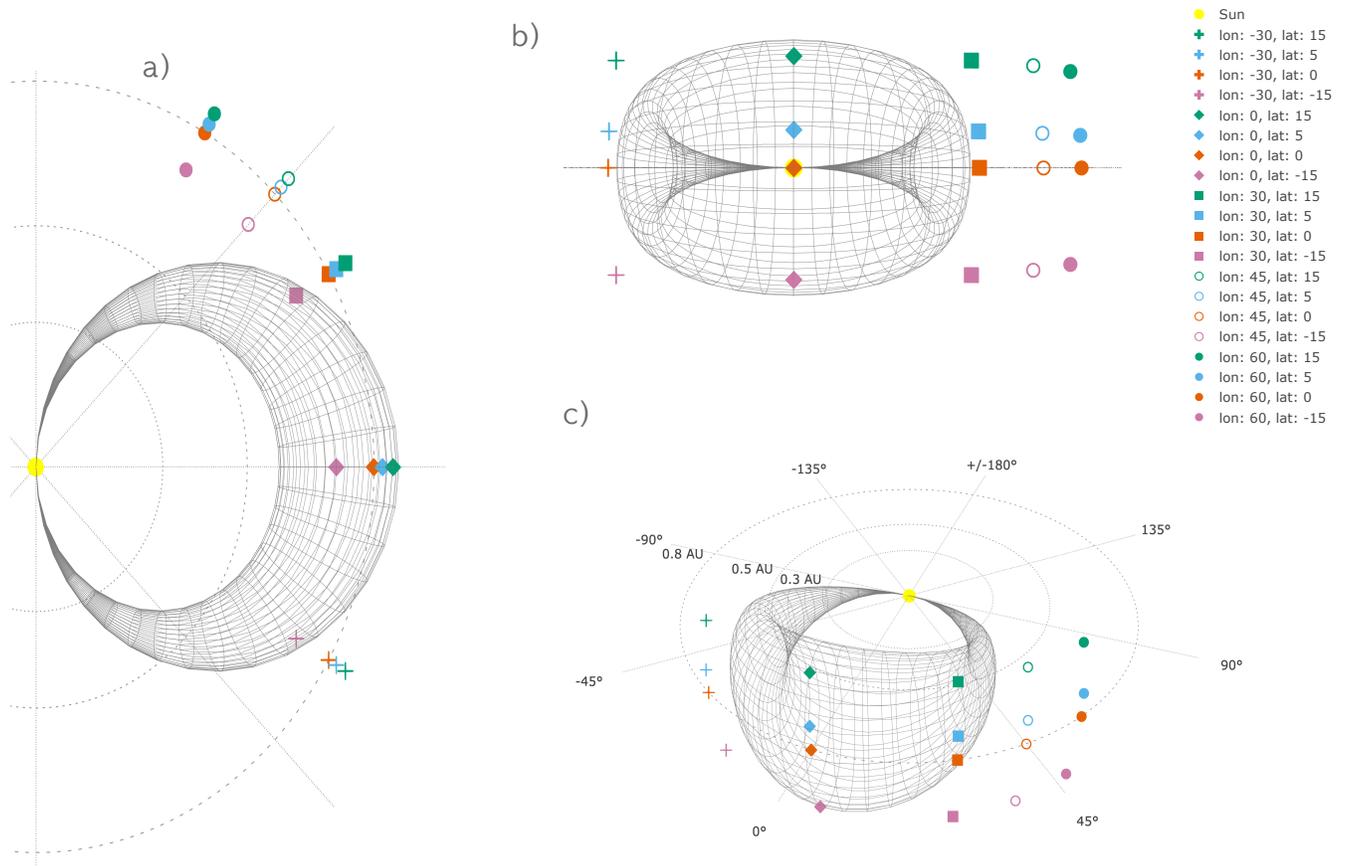}}
\caption{Synthetic spacecraft setup for the low inclination flux rope, shown from three different viewing angles: a) top view down from solar equatorial north, b) front view from the apex of the structure along the solar equatorial plane, c) angled view. Spacecraft positions include five longitudinal (the central meridian with respect to the CME, $\pm 30\dgg$, $+ 45\dgg$, and $+ 60\dgg$) and four latitudinal planes (at the solar equatorial plane, $\pm 5 \dgg$, and $+ 15 \dgg$) at 0.8~au radial distance from the Sun.} 
\label{fig:setup}
\end{figure*}

Figure \ref{fig:setup} presents an overview of the experiment setup for a low inclination flux rope, placing 20 synthetic spacecraft at a radial distance of 0.8~au from the Sun. Due to the simplified propagation and lack of interaction within the model, we focus on a single radial distance in our study. This approach is supported by \cite{Scolini_2023}, who reported a lack of increase in the magnetic complexity of CMEs in the absence of interactions with other large-scale structures. Furthermore, due to the self-similar expansion of the model, observations at 0.8~au are representative of those at L1. Placing the spacecraft at this distance avoids the added complexity of differing arrival times at various longitudes, which would complicate interpretation of results for spacecraft positioned at a greater distance from the Sun. While the intermediate radial distance of 0.8 au is not as close to the Sun as the perihelia of Solar Orbiter or Parker Solar Probe, it still provides a realistic setting for typical events measured by spacecraft in the inner heliosphere. The chosen distance ensures that in situ signatures span a sufficiently long time period to enable adequate interpretability of results and comparison between spacecraft. These considerations yield an experimental setup that effectively balances the need for representative data and manageable analysis.
The respective positions of the spacecraft include five longitudinal (the central meridian with respect to the CME, $\pm 30\dgg$, $+45\dgg$ and $+ 60\dgg$) and four latitudinal planes (at the solar equatorial plane, $\pm 5 \dgg$, and $+ 15 \dgg$). These are visualized by different colors representing different latitudinal planes, as well as different symbols for different longitudinal planes. The asymmetry of the spacecraft distribution was chosen due to the inherent symmetry of the structure introducing redundancy in the results.

\begin{figure*}[h!]
\centering
{\includegraphics[width=\textwidth]{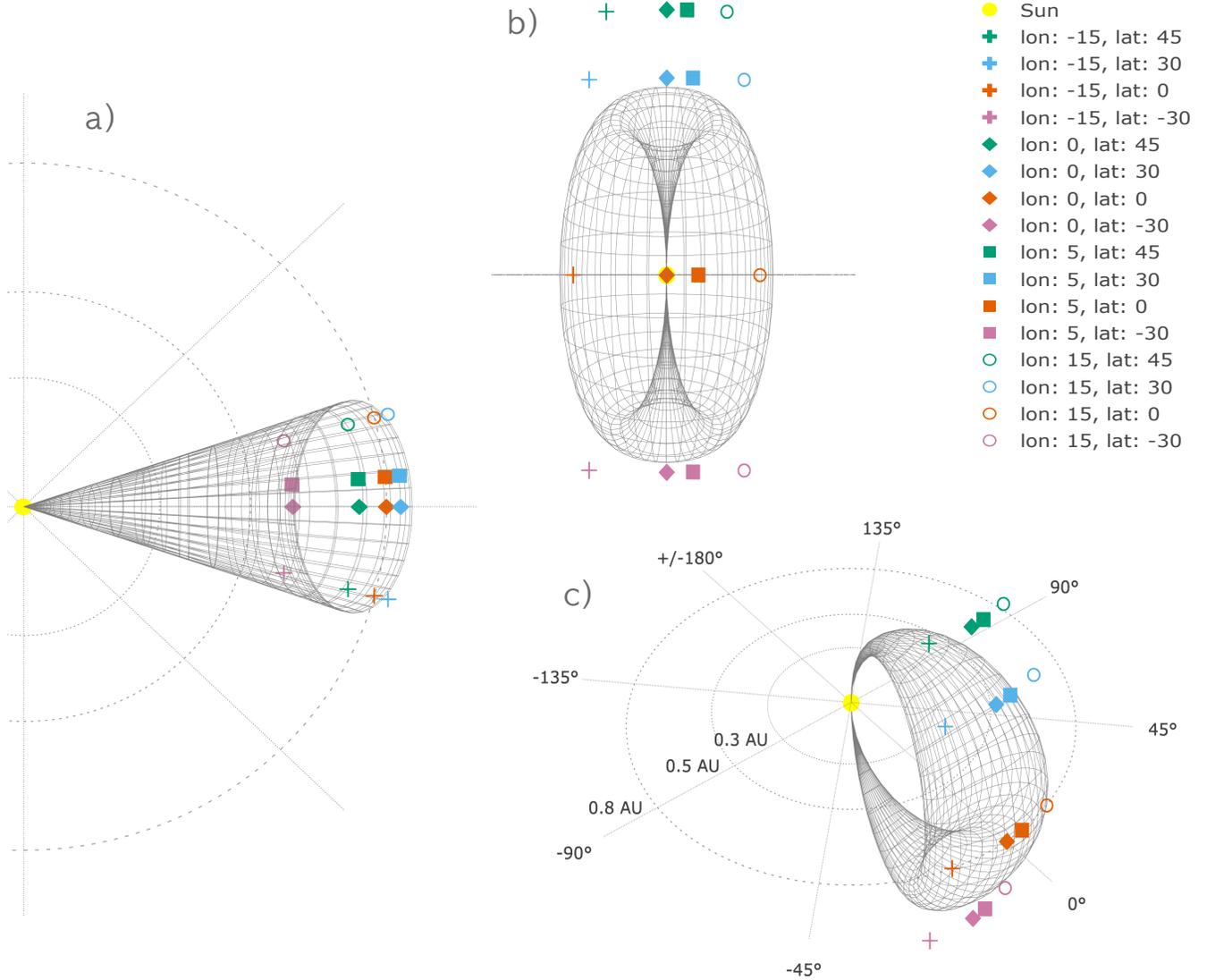}}
\caption{Synthetic spacecraft setup for the high inclination flux rope, shown from three different viewing angles: a) top view down from solar equatorial north, b) front view from the apex of the structure along the solar equatorial plane, c) angled view. Spacecraft positions include four longitudinal (the central meridian with respect to the CME, $\pm 5\dgg$, and $+ 15\dgg$) and four latitudinal planes (at the solar equatorial plane, $\pm 30 \dgg$, and $+ 45 \dgg$) at 0.8~au radial distance from the Sun.} 
\label{fig:setuphigh}
\end{figure*}

Figure \ref{fig:setuphigh} shows the simulation setup similarly for a high inclination flux rope, with 16 spacecraft in four longitudinal (the central meridian with respect to the CME, $\pm 5\dgg$ and $+ 15\dgg$) and four latitudinal planes (at the solar equatorial plane, $\pm 30 \dgg$, and $+ 45 \dgg$), also at a radial distance of 0.8~au from the Sun. Since the longitudinal extension of the high inclination flux rope is much smaller than for the low inclination flux rope, fewer longitudinal planes are included, while the four latitudinal planes are stretched out to account for the higher latitudinal extension in this case.

\begin{figure*}[h!]
\centering
{\includegraphics[width=\textwidth]{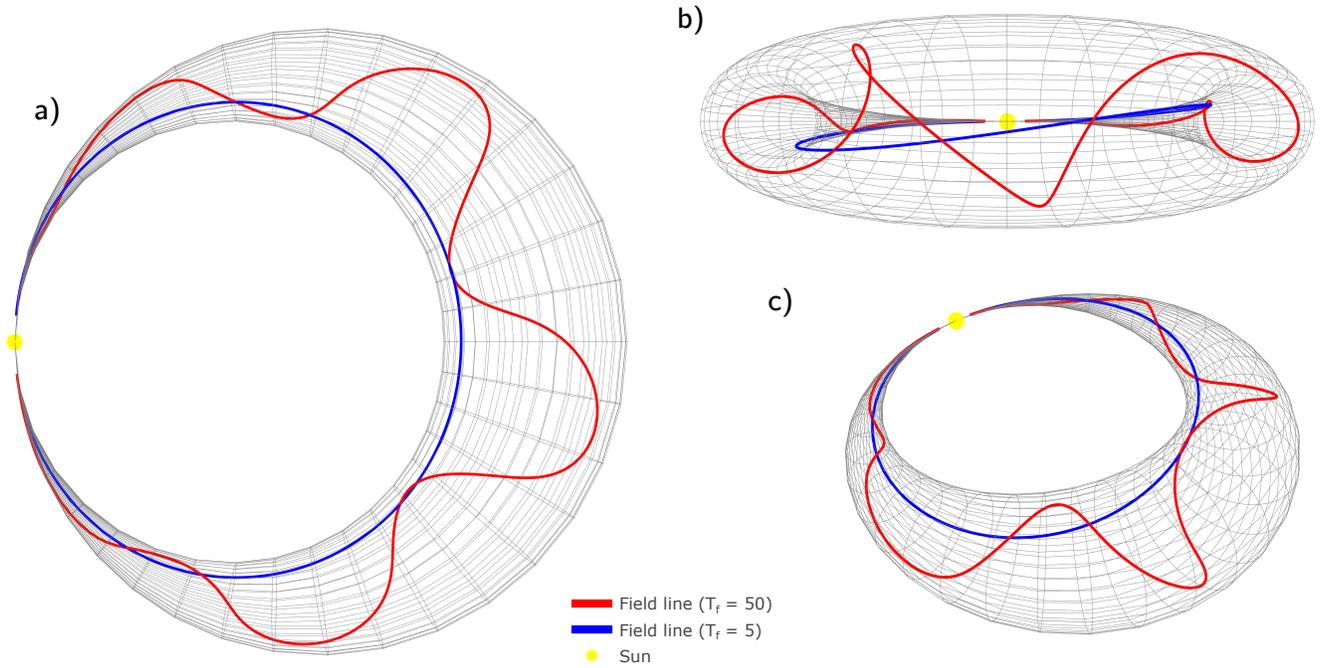}}
\caption{A visualization of two field lines within the CME shape for high twist of $T_f = 50$ (red) and low twist of $T_f = 5$ (blue).} 
\label{fig:fieldlines}
\end{figure*}

The initial parameters as used for the model in 3DCOREweb are shown in Table \ref{tab:3dcore_params}. While inclination and twist are varied accordingly to produce the eight different flux rope types (see Table \ref{tab:fluxtypes}), the other parameters are kept to generic standard values to better compare the influence of the observers' locations. Due to the model setup, inclination values of $0 \dgg$ and $180 \dgg$ correspond to low inclination flux ropes with their central axis pointing in opposite directions. Inclination values of $90 \dgg$ and $270 \dgg$ lead to both orientations of high inclination flux ropes. For the flux rope example in this study, a $T_{f}=50$ corresponds to a high twist number of $\tau \approx5$ turns and a $T_{f}=5$ corresponds to a low twist number of $\tau \approx0.5$ turns. The twist number $\tau$ is defined as the total number of twists along the entire torus. These numbers are chosen in agreement with \cite{Kahler_2011_length}, who estimated the number of field line rotations over the assumed CME full length to be 1--10 turns. A visualization of a single field line for both low and high twist CME can be seen in Figure \ref{fig:fieldlines}, where the low twist field line is represented in blue, and the high twist in red.

\begin{table}
\centering
\caption{Model parameters longitude, latitude, inclination, diameter at 1 au ($D_{1\mathrm{au}}$), aspect ratio ($\delta$), launch radius ($R_{0}$), launch velocity ($V_{0}$), expansion rate ($n_{a}$), background drag ($\Gamma$), solar wind speed ($V_{sw}$), twist factor ($T_{f}$), magnetic decay ($n_b$), magnetic field strength at 1 au ($B_{1\mathrm{AU}}$), for low inclination and high inclination flux ropes with low and high $T_{f}$. The longitude and latitude of the flux rope apex are given in HEEQ coordinates.}
\begin{tabular}{l c r}
Parameters & Values & Units \\[0.5ex]
\hline
Longitude & $0.$ & deg \\[0.5ex]
Latitude & $0.$ & deg \\[0.5ex]
Inclination & $0.$ / $90.$ / $180.$ / $270.$ & deg \\[0.5ex]
$D_{1\mathrm{AU}}$ & $0.3$ & au \\[0.5ex]
$\delta$ & $2.$ &  \\[0.5ex]
$R_{0}$ & $20.$ & solar radii \\[0.5ex]
$V_{0}$ & $600.$ & km/s \\[0.5ex]
$n_{a}$ & $1.14$ &  \\[0.5ex]
$\Gamma$ & $1.$ &  \\[0.5ex]
$V_{\mathrm{sw}}$ & $500.$ & km/s \\[0.5ex]
$T_{f}$ &  $50.$ / $5.$ / $-5.$ / $-50.$ & \\[0.5ex]
$n_{b}$ &  $1.64$ & \\[0.5ex]
$B_{1\mathrm{au}}$ & $25.$ & nT \\[0.5ex]
\hline
\end{tabular}
\label{tab:3dcore_params}
\end{table}

\begin{table}
\centering
\caption{Inclination and $T_{f}$ as set in the 3DCORE model for different flux rope types including both low and high twist values.}
\begin{tabular}{l c c c c}
Flux rope type & Handedness & Inclination [deg] & $T_{f}$ (high twist) & $T_{f}$ (low twist) \\
\hline
SWN & Right-handed & $180.$ & $50.$ & $5.$ \\
NES & Right-handed & $0.$ & $50.$ & $5.$ \\
NWS & Left-handed &  $180.$ & $-50.$ & $-5.$ \\
SEN & Left-handed &  $0.$ & $-50.$ & $-5.$ \\
WNE & Right-handed & $270.$ &  $50.$ & $5.$ \\
ESW & Right-handed & $90.$ &  $50.$ & $5.$ \\
ENW & Left-handed & $270.$ &  $-50.$ & $-5.$ \\
WSE & Left-handed & $90.$ &  $-50.$ & $-5.$ \\
\hline
\end{tabular}
\label{tab:fluxtypes}
\end{table}

By comparing the resulting in situ profiles, we investigate which circumstances lead to CNRs. More precisely, we analyse the influence of flux rope type, twist parameter and observer location on the measured in situ signatures. Furthermore, we attempt to show whether flux rope types can be unambiguously distinguished.

\section{Comparison of Signatures} \label{sec:results}

Figures \ref{fig:low_inc_high_twist_SWN_low} -- \ref{fig:high_inc_low_twist_WNE} show the results. Each figure presents the synthetic in situ measurements for a specific flux rope type as measured by the observers described in Section \ref{sec:experimentsetup}. Each subplot coincides with one spacecraft, with each row representing a different latitude and each column representing a different longitude. The vertical axis shows the magnetic field $B(t)$ in nT and the horizontal axis indicates the time that has passed since the launch of the CME in hours. All three magnetic field components (radial $B_{r}$, tangential $B_{t}$, normal $B_{n}$), as well as the total magnetic field ($B_{tot}$) are shown in RTN coordinates. We have chosen one low inclination and one high inclination flux rope type, each with low twist and high twist configurations. The general tendencies of how the signatures change upon varying longitude and latitude are the same for each low/high flux rope type and only differ regarding its handedness. Nevertheless, these figures have been created for all flux rope types for completeness and can be found in the Appendix.

\subsection{High Twist Flux Ropes}

\begin{figure*}[h!]
\centering
{\includegraphics[width=\textwidth]{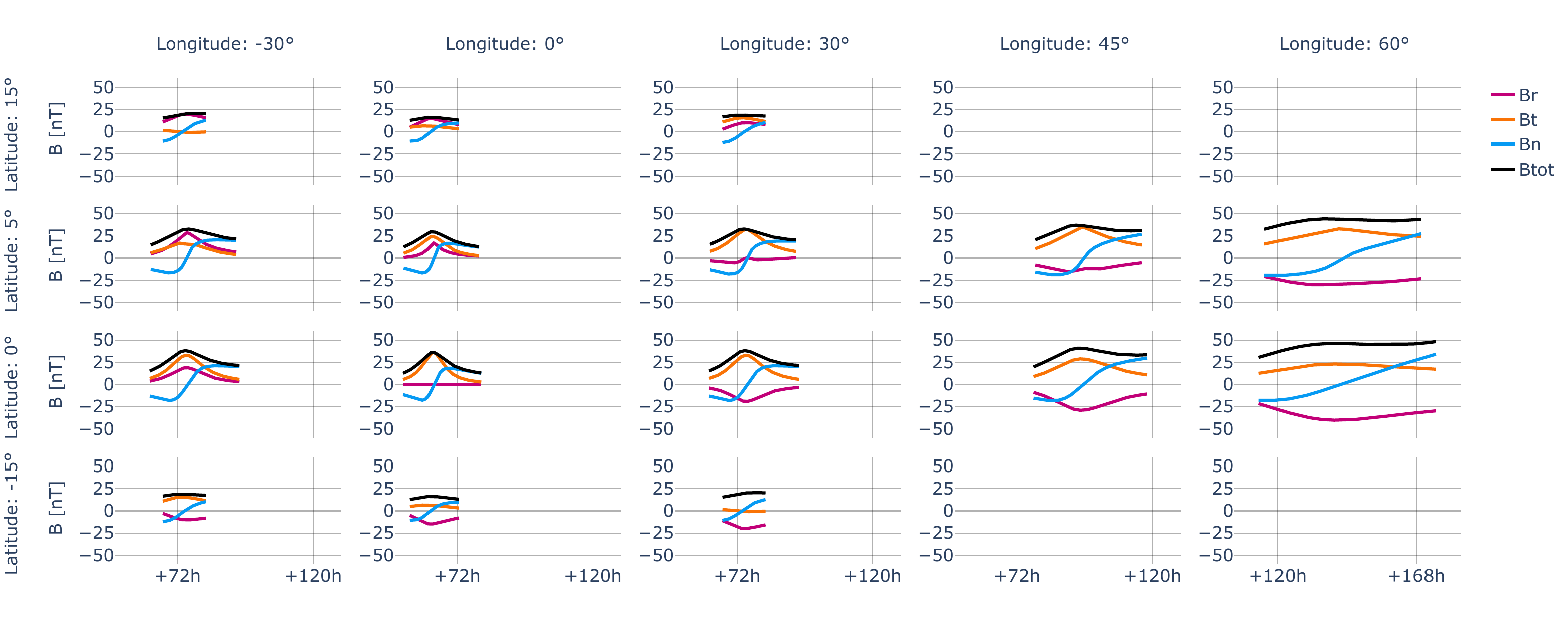}}
\caption{In situ measurements of the synthetic spacecraft in different latitudes and longitudes for a low inclination and high twist SWN type flux rope. For better comparison of the times of arrival, the x-axis displays the same time range for all subplots except for a longitude of $60\dgg$.} 
\label{fig:low_inc_high_twist_SWN_low}
\end{figure*}

Figure \ref{fig:low_inc_high_twist_SWN_low} shows the results for a low inclination and high twist SWN flux rope type. As expected, the in situ signatures exhibit a bipolar $B_{n}$ component, changing sign from negative (South) to positive (North), and a unipolar $B_{t}$ component that stays positive (West). 

Varying the spacecraft longitude from $-30\dgg$ to $+60\dgg$, $B_{r}$ shifts from positive to negative. Similar behaviour can also be observed when varying latitude, with a negative latitude corresponding to a negative $B_{r}$ component and a positive latitude corresponding to a positive $B_{r}$. The $B_{r}$ component stays 0 for the apex hit at both longitude and latitude equal to 0. This pattern is also observed in other classic flux rope models such as the Lundquist model, where the deviation of $B_{r}$ away from 0 gives a sense of the distance of the spacecraft from the flux rope axis, otherwise known as the impact parameter \citep{good2018correlation, lepping1990magnetic, lepping2006}. Additionally, moving from the apex to the flanks of the CME, the profile gets stretched in time and the total change of the values in all components becomes smaller. For the flank encounters, high latitudes do not measure any signatures as the CME does not hit a synthetic spacecraft at that point, as expected. Additionally, a larger radial width of the CME is traversed for a latitude of $0\dgg$ compared to high latitudes.  The high twist of the flux rope is reflected in the significant rotation of the magnetic field vector leading to a variation in all three components for apex hits and still a considerable variation in $B_{n}$ for flank encounters. However, the overall profile is notably stretched and flattened, meaning a reduced variation in each magnetic field component. In future work, this could be investigated more closely by calculating the difference between the maximum and minimum value of each component. As this lies beyond the scope of our current study, we determine the flatness of a profile only through visual inspection. In the extreme cases of a latitude of $- 15\dgg$ combined with a longitude of $30\dgg$ or a latitude of $ 15\dgg$ combined with a longitude of $-30\dgg$, $B_{t}$ is observed very close to 0. 

Interestingly, the total magnetic field increases significantly for flank hits. For an apex hit, the maximum $B_{tot}$ is only slightly more than 25 nT, while for a longitude of $60\dgg$, $B_{tot}$ almost reaches 50 nT. The observed effect arises from the tapering of the torus shape, resulting in an inverse relationship between the torus radius and magnetic field strength. Nevertheless, the validity of this assumption may be refined through future investigations into multipoint events.

\begin{figure*}[h!]
\centering
{\includegraphics[width=\textwidth]{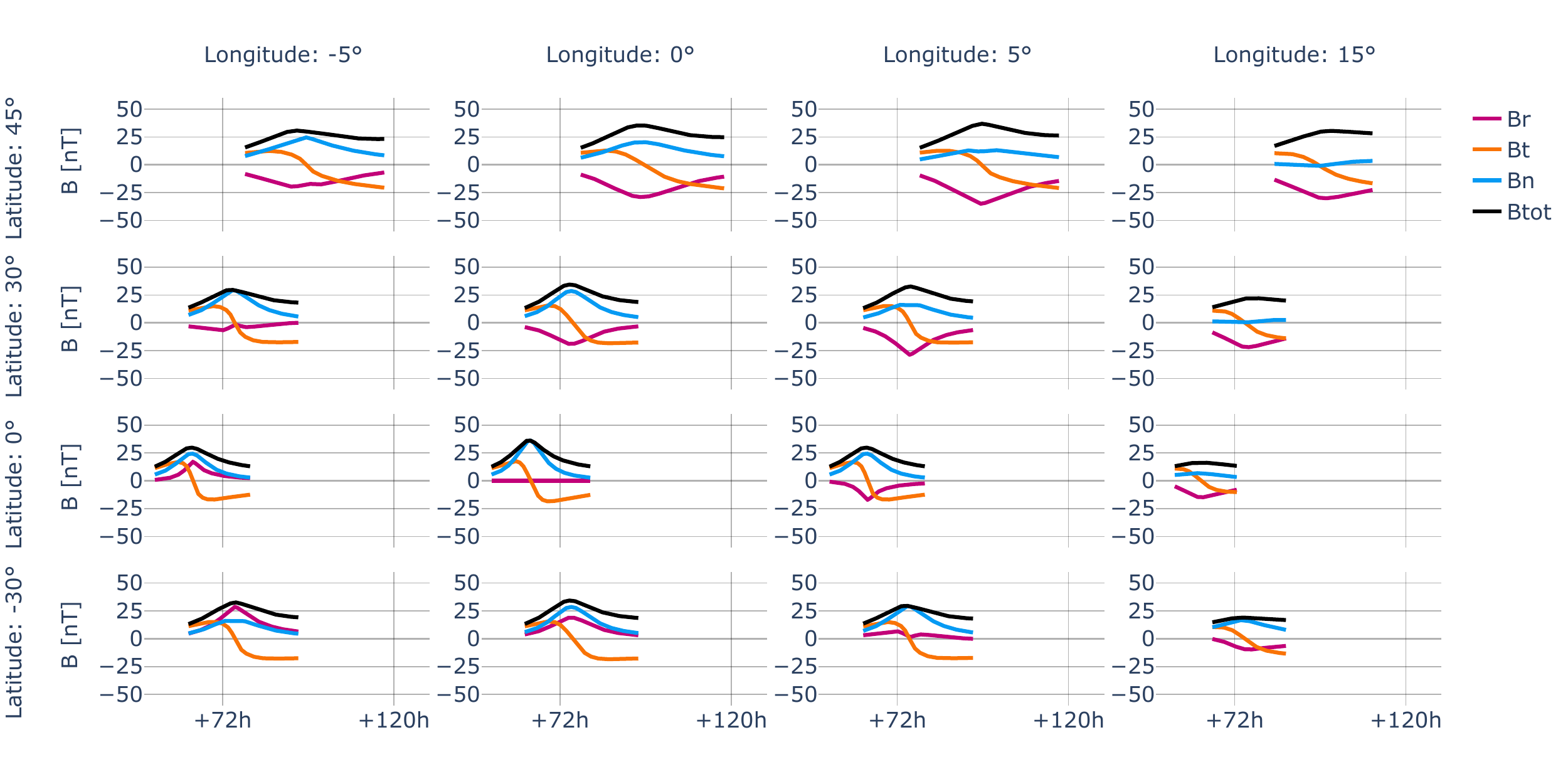}}
\caption{In situ measurements of the synthetic spacecraft in different latitudes and longitudes for a high inclination and high twist WNE type flux rope. For better comparison of the times of arrival, the x-axis displays the same time range for all subplots.} 
\label{fig:high_inc_high_twist_WNE}
\end{figure*}

Figure \ref{fig:high_inc_high_twist_WNE} shows a high inclination and high twist WNE flux rope type. The high inclination of the flux rope can be seen in the bipolarity of $B_{t}$, changing sign from positive (West) to negative (East), and the unipolarity of $B_{n}$, staying positive (North).

In the high inclination case in general, high latitudes correspond to flank encounters, while the spacecraft in the solar equatorial plane experience apex hits. Keeping this in mind, we observe similar patterns as for the high twist, low inclination flux rope types (for example Figure \ref{fig:low_inc_high_twist_SWN_low}), with flank encounters producing considerably flatter in situ profiles compared to the apex hits.

Interestingly, we observe no CNRs for any of the high twist flux rope types, even for flank hits. While the unipolar components show a less significant variation for flank hits, the bipolarity of the respective component is conserved throughout the whole structure. As a result, the components analysed for the determination of the flux rope types are unambiguous, regardless of where the CME hits the spacecraft. Nevertheless, we observe some extreme cases for which the unipolar component is observed very close to 0 which might lead to a misclassification of the flux rope type. One such example would be the crossing at a longitude of $-30\dgg$ and a latitude of $15\dgg$ for the SWN type shown in Figure \ref{fig:low_inc_high_twist_SWN_low}. Since the tangential component is very close to 0, the flux rope might be mistaken as SEN type, as shown in Figure \ref{fig:low_inc_high_twist_SEN_low} of the appendix for the crossing at a longitude of $30\dgg$ and a latitude of $15\dgg$.

\subsection{Low Twist Flux Ropes}

\begin{figure*}[h!]
\centering
{\includegraphics[width=\textwidth]{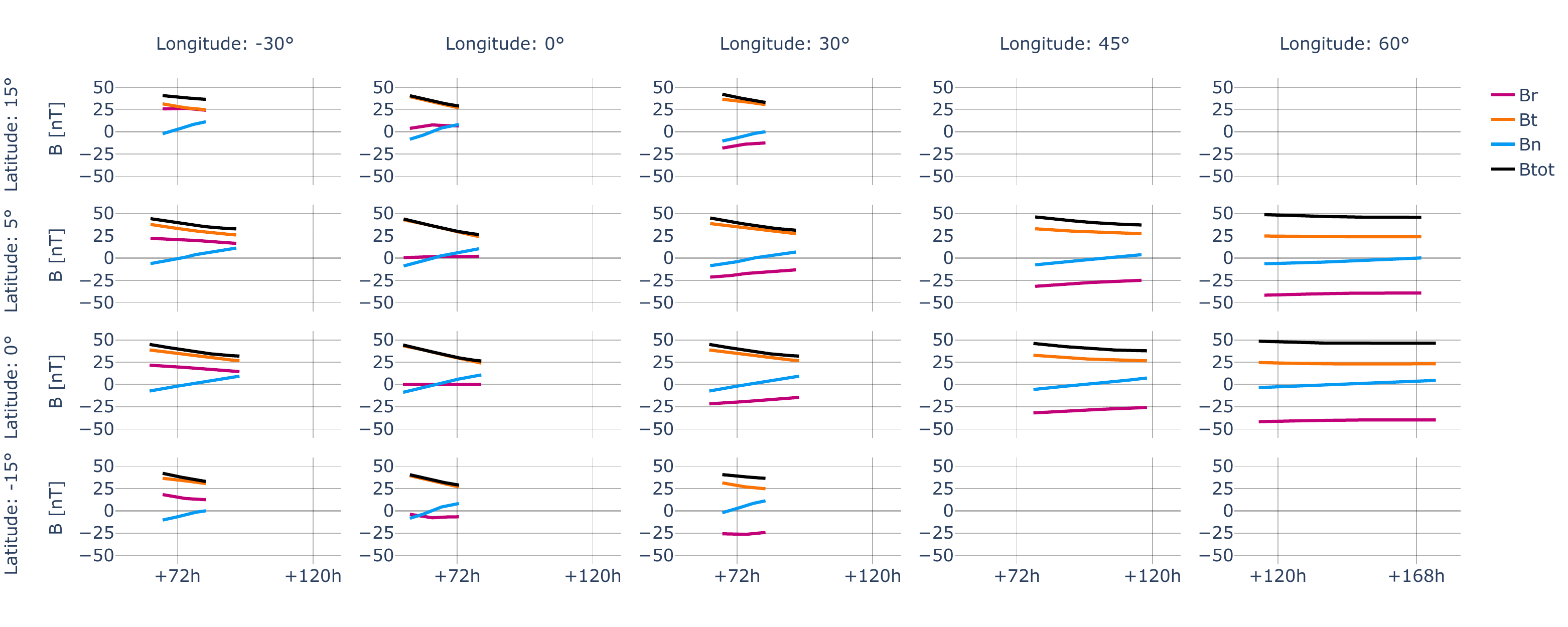}}
\caption{In situ measurements of the synthetic spacecraft in different latitudes and longitudes for a low inclination and low twist SWN type flux rope. For better comparison of the times of arrival, the x-axis displays the same time range for all subplots except for a longitude of $60\dgg$.} 
\label{fig:low_inc_low_twist_SWN_low}
\end{figure*}

\begin{figure*}[h!]
\centering
{\includegraphics[width=\textwidth]{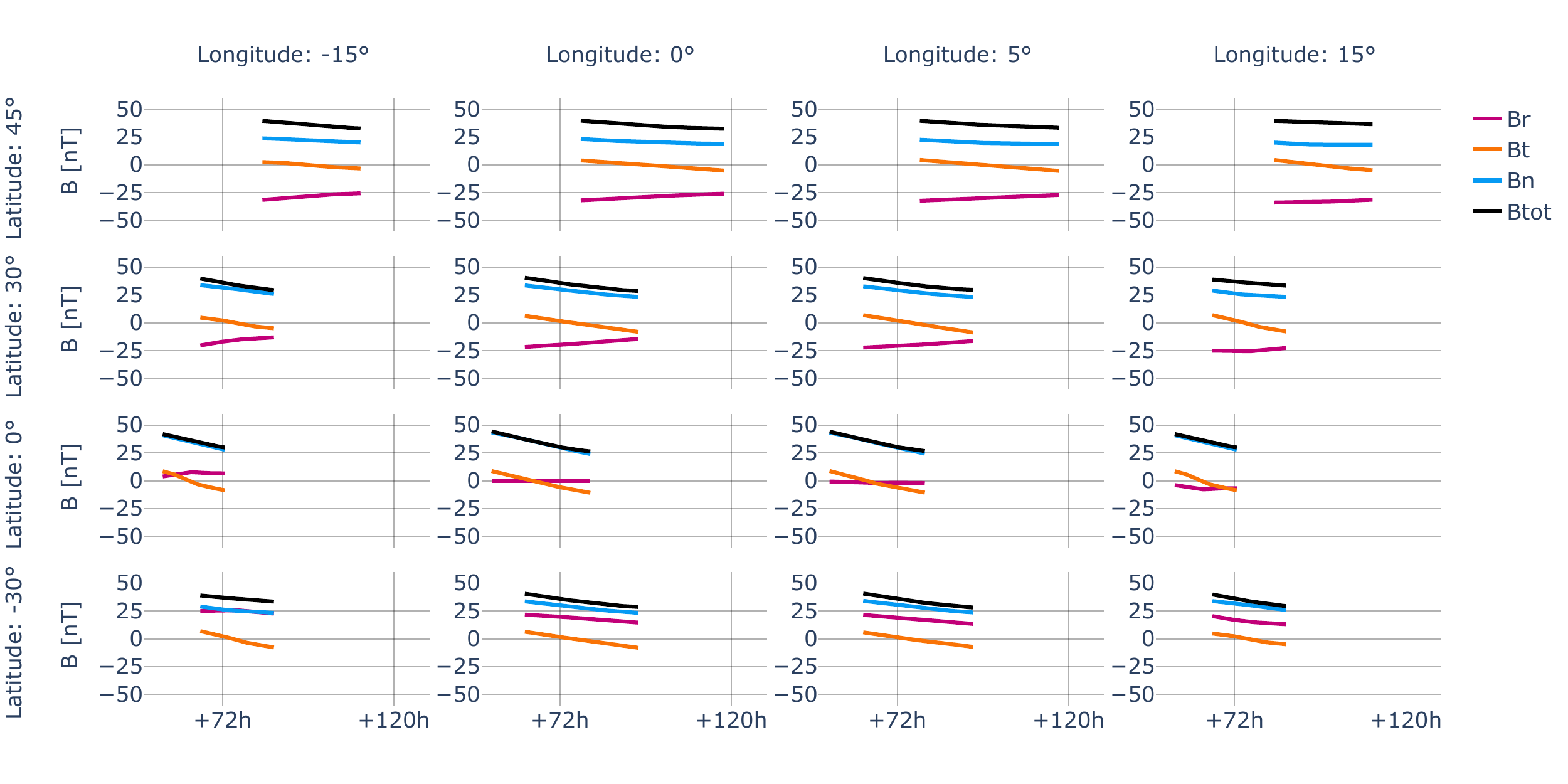}}
\caption{In situ measurements of the synthetic spacecraft in different latitudes and longitudes for a high inclination and low twist WNE type flux rope. For better comparison of the times of arrival, the x-axis displays the same time range for all subplots.} 
\label{fig:high_inc_low_twist_WNE}
\end{figure*}

Figures \ref{fig:low_inc_low_twist_SWN_low} and \ref{fig:high_inc_low_twist_WNE} show the low twist counterparts to Figures \ref{fig:low_inc_high_twist_SWN_low} and \ref{fig:high_inc_high_twist_WNE}. Figure \ref{fig:low_inc_low_twist_SWN_low} shows the results for a low inclination and low twist SWN flux rope type. The low twist of the flux rope leads to a very minor rotation of the magnetic field. Compared to the pronounced bipolarity of $B_{n}$ for the high twist SWN flux rope, only a very minor change from slightly negative to slightly positive can be observed in this plot. Nevertheless, as $B_{t}$ appears very flat, it is constantly positive and has a greater difference to 0 than for the high twist counter part.

While apex hits already show very flat signatures, we observe CNRs for flank hits. In contrast to Figure \ref{fig:low_inc_high_twist_SWN_low} and \ref{fig:high_inc_high_twist_WNE}, that showed significantly higher $B_{tot}$ values for flank hits compared to the corresponding apex hits, the maximum total magnetic field is only slightly higher for flank hits compared to an apex hit in this example.

Figure \ref{fig:high_inc_low_twist_WNE} shows a high inclination and low twist WNE flux rope type. The high inclination of the flux rope can once again be identified by the bipolarity of $B_{t}$, changing sign from positive to negative, and the unipolarity of $B_{n}$, staying positive. Nevertheless, as for the low twist SWN type (Figure \ref{fig:low_inc_low_twist_SWN_low}), the observed bipolarity is very minor in the WNE type (Figure \ref{fig:high_inc_low_twist_WNE}).

As for all low inclination, low twist types, flank encounters produce considerably flatter in situ profiles compared to the apex hits, leading to CNR signatures. This makes a distinction of flux rope types based on in situ signatures difficult, even for perfectly radial encounters. However, it can be observed that several cases are almost indistinguishable (for example Figure \ref{fig:high_inc_low_twist_WNE}, latitude $45\dgg$, longitude $-15\dgg$ and $15\dgg$). When disregarding the duration of the signatures, the difference between cases is even more negligible. For example in Figure \ref{fig:high_inc_low_twist_WNE}, at a latitude of $45\dgg$ crossings at all depicted longitudes ($-15\dgg$,  $0\dgg$, $5\dgg$ and $15\dgg$) produce qualitatively equal results only varying in their temporal extension. Given that this is a parametric study with known CME locations in the heliosphere, we can distinguish between different profiles of flank encounters. However, in a real scenario, distinguishing between two similar profiles, such as in Figure \ref{fig:high_inc_low_twist_WNE} at $45\dgg$ latitude and longitudes of $-15\dgg$ and $15\dgg$, would be challenging without knowing the global geometry. This highlights that the identification of the flux rope type and encounter is not completely unambiguous, especially for flank hits that produce similar in situ signatures.

\section{Discussion and Conclusions} \label{sec:discussion}

In this study, we investigated the effect of the position of a synthetic spacecraft and the resulting trajectory at different latitudinal and longitudinal locations at 0.8~au on the in situ magnetic field signatures observed using 3DCOREweb. We investigated hypothesis (1), whether trajectory effects could be a possible explanation for the observation of CNRs. We discover that for a high twist CME, a flank hit results in a very flat in situ profile with two out of three components not exhibiting a visible variation that would indicate a rotation of the magnetic field vector. Nevertheless, the third component still displays a bipolar signature, although it is much flatter than observed for an apex hit. To observe a CNR, the CME had to be initialized with a low twist and a flank hit had to be simulated. Relating our findings to the general context of whether the legs of a CME might contain twisted field lines, as studied in \cite{owens2016legs}, further investigation of multipoint events will be needed. While our initial findings suggest that observing a constant field for a flank impact implies minimal twisting within its legs, confirming this requires simultaneous measurements at the apex. This would solidify whether the observed CME's structure has varying levels of twist throughout, rather than assuming it to be a generally low twist CME.

This also points out an obvious drawback of our model, namely the constant twist and its non-deformable shape. The opening angle of $180\dgg$ prevents the modeling of a trajectory that experiences both an apex hit and a flank encounter for the same CME. Keeping in mind that CMEs are thought to be generally deformable structures \citep[e.g.][]{manchester2017physical, kilpua2019, luhmann2020}, which may deviate strongly from our idealized shape and thus in reality quite often experience such scenarios, limits the accuracy of our model. Nevertheless, the separate analysis of flank encounters and apex hits ties in with the possible explanation of trajectory effects causing CNR signatures.

In the past, intrinsic flux rope types, as identified in solar signatures, have often not matched the in situ flux rope types identified as investigated in e.g. \cite{palmerio2018coronal}. Throughout this work, we compared in situ results for different flux rope types and investigated hypothesis (2), whether flux rope types could be misinterpreted due to non-apex hits. For a low twist CME, it appeared indeed to be challenging to identify the flux rope type, due to CNRs observed for flank hits. Nevertheless, for high twist CMEs, the flux rope type could be unambiguously assigned, regardless of where the impact occurred, unless one of the identified rare cases of an almost zero unipolar component occurred. While this hints towards the need for alternative explanations for the mismatching of flux rope types from solar to in situ observations, we have to consider that our model poses some very general assumptions. For a deformable CME, as well as non-frontal encounters, this could nonetheless be explained by trajectories and needs to be further investigated. For a highly deformed flux rope, a spacecraft could potentially be impacted by the front part of a CME, followed directly by a flank encounter. In comparison to that, the self-similar expansion implemented in our model combined with a stationary spacecraft leads to a highly idealized frontal encounter with the flux rope, which might only partially model reality. Especially for spacecraft in close proximity to the Sun, such as Parker Solar Probe, a rapidly changing spacecraft position may have to be taken into account when analysing CME encounters. Even though this has been studied in \cite{moestl2020prediction} and the effect turned out to be negligible, this study only looked at one particular type of encounter.

To further corroborate the validity of our results, we are awaiting the possible occurrence of many more multipoint measurements in the near future. Especially with Solar Orbiter due to commence probing higher latitudes and the general increase in spacecraft measuring the solar wind, including the future operational space weather mission Vigil, planned for launch by ESA in 2031, as well as the approach of the maximum of solar cycle 25, statistical studies will hopefully be facilitated.

Throughout this work, we confirm the substantial impact of both latitudinal and longitudinal positions of spacecraft on the appearance of in situ signatures. These results could significantly aid in future analysis of in situ data, particularly in concluding the relative trajectory of the spacecraft through the CME with respect to its propagation direction. However, it is crucial to acknowledge the potential challenges in discerning between similar profiles, especially in real-life scenarios. Our research highlights that, especially in cases of low twist, distinctions among signatures at different locations may be subtle, often distinguishable solely by their duration. Yet, in real-world contexts, the duration could as well be attributed to the CMEs thickness, complicating the determination of spacecraft trajectory based solely on in situ profiles. Nonetheless, integrating additional data sources, such as remote sensing observations or complementary spacecraft measurements, holds promise for resolving certain ambiguities and deriving meaningful conclusions regarding spacecraft trajectory.

In summary, our work not only contributes to advancing our understanding of in situ observations but also underscores the necessity for a comprehensive approach in interpreting such data, ultimately enhancing our ability to navigate the complexities of space weather phenomena.

\begin{acknowledgments}

For the development of the 3DCOREweb application, we have benefited from the availability of SolO, STEREO, Wind, PSP, and BepiColombo data, and thus would like to
thank the instrument teams and data archives for their data distribution efforts. H.T.~R., U.V.~A., C.~M., and E.~D. are funded by the European Union (ERC, HELIO4CAST, 101042188). 
Views and opinions expressed are however those of the author(s) only and do not necessarily reflect those of the European Union or the European Research Council Executive Agency. Neither the European Union nor the granting authority can be held responsible for them.
A.J.~W. acknowledges the financial
support by an appointment to the NASA Postdoctoral Program at NASA Goddard Space Flight Center, administered by Oak Ridge Associated Universities through a contract with NASA. J.~L. acknowledges that this research was funded in whole, or in part, by the Austrian Science Fund (FWF) [P 36093]. For open access purposes, the author has applied a CC BY public copyright license to any author-accepted manuscript version arising from this submission. A living version of 3DCOREweb can be found at \url{https://github.com/hruedisser/3DCOREweb}. To allow the community to compare future studies with our findings, the source code to generate the figures is available as a Jupyter notebook at \url{https://github.com/hruedisser/3DCOREweb/blob/main/Signatures_Paper.ipynb} and is preserved on Zenodo at \url{https://doi.org/10.5281/zenodo.11445313}. Finally, we sincerely thank the editor and reviewer for their helpful comments and suggestions to substantially improve the quality of our research.

\end{acknowledgments}

\newpage
\appendix

\begin{figure}[ht!]
\centering
{\includegraphics[width=\textwidth]{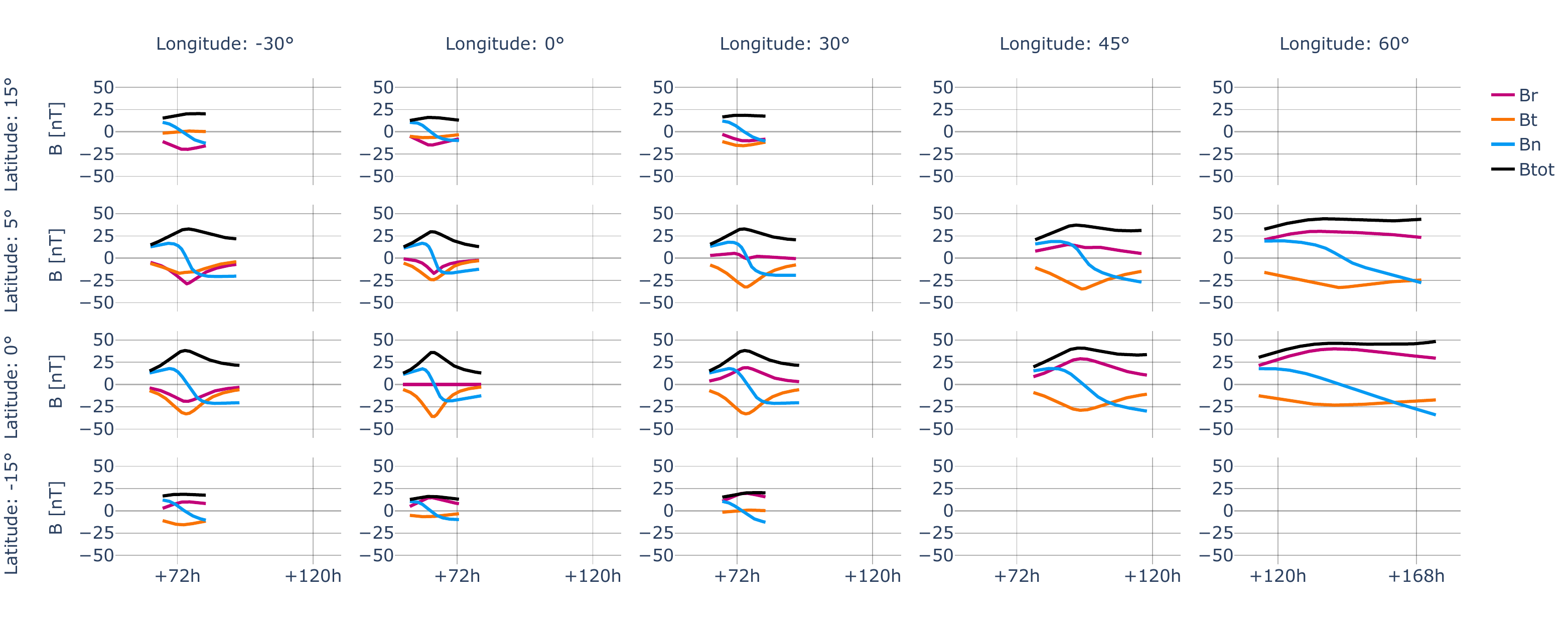}}
\caption{In situ measurements of the synthetic spacecraft in different latitudes and longitudes for a low inclination and high twist NES type flux rope. For better comparison of the times of arrival, the x-axis displays the same time range for all subplots except for a longitude of $60\dgg$.} 
\label{fig:low_inc_high_twist_NES_low}
\end{figure}

The second right-handed low inclination flux rope type NES is shown in Figure \ref{fig:low_inc_high_twist_NES_low}, for which $B_{n}$ changes sign from positive to negative and  $B_{t}$ remains negative.


\newpage

\begin{figure*}[ht!]
\centering
{\includegraphics[width=\textwidth]{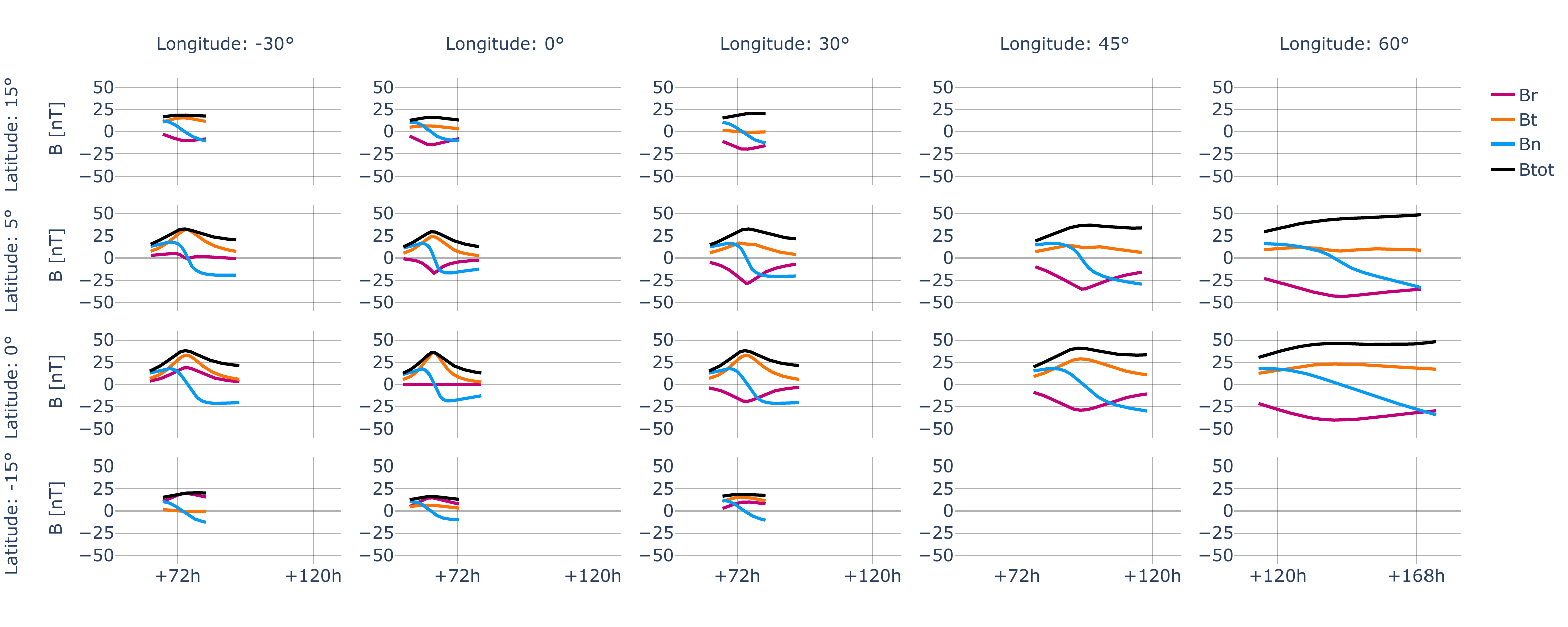}}
\caption{In situ measurements of the synthetic spacecraft in different latitudes and longitudes for a low inclination and high twist NWS type flux rope. For better comparison of the times of arrival, the x-axis displays the same time range for all subplots except for a longitude of $60\dgg$.} 
\label{fig:low_inc_high_twist_NWS_low}
\end{figure*}

\begin{figure*}[h!]
\centering
{\includegraphics[width=\textwidth]{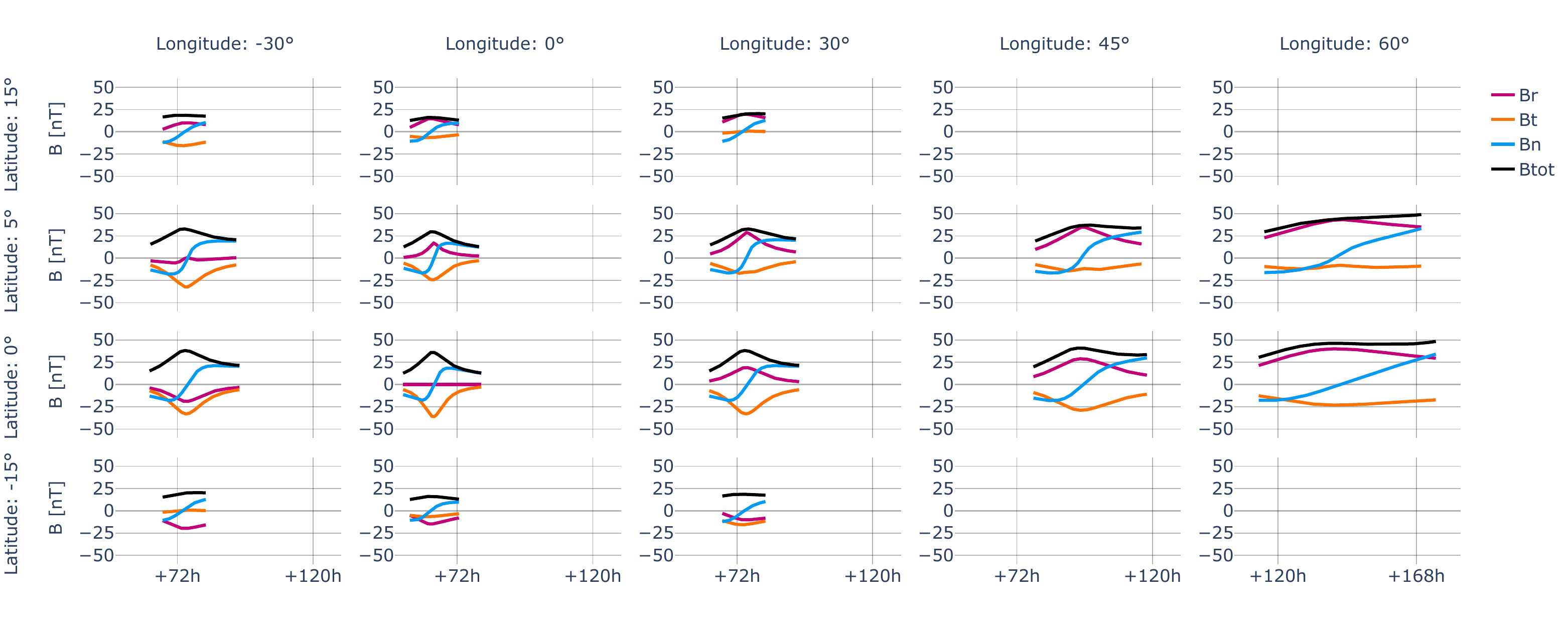}}
\caption{In situ measurements of the synthetic spacecraft in different latitudes and longitudes for a low inclination and high twist SEN type flux rope. For better comparison of the times of arrival, the x-axis displays the same time range for all subplots except for a longitude of $60\dgg$.} 
\label{fig:low_inc_high_twist_SEN_low}
\end{figure*}

Figures \ref{fig:low_inc_high_twist_NWS_low} and \ref{fig:low_inc_high_twist_SEN_low} show the left-handed counterparts NWS and SEN to these, where the NWS type exhibits $B_{n}$ changing sign from positive to negative and  $B_{t}$ remaining positive, while the SEN type has $B_{n}$ changing sign from negative to positive and  $B_{t}$ remaining negative.

\newpage

\begin{figure*}[h!]
\centering
{\includegraphics[width=\textwidth]{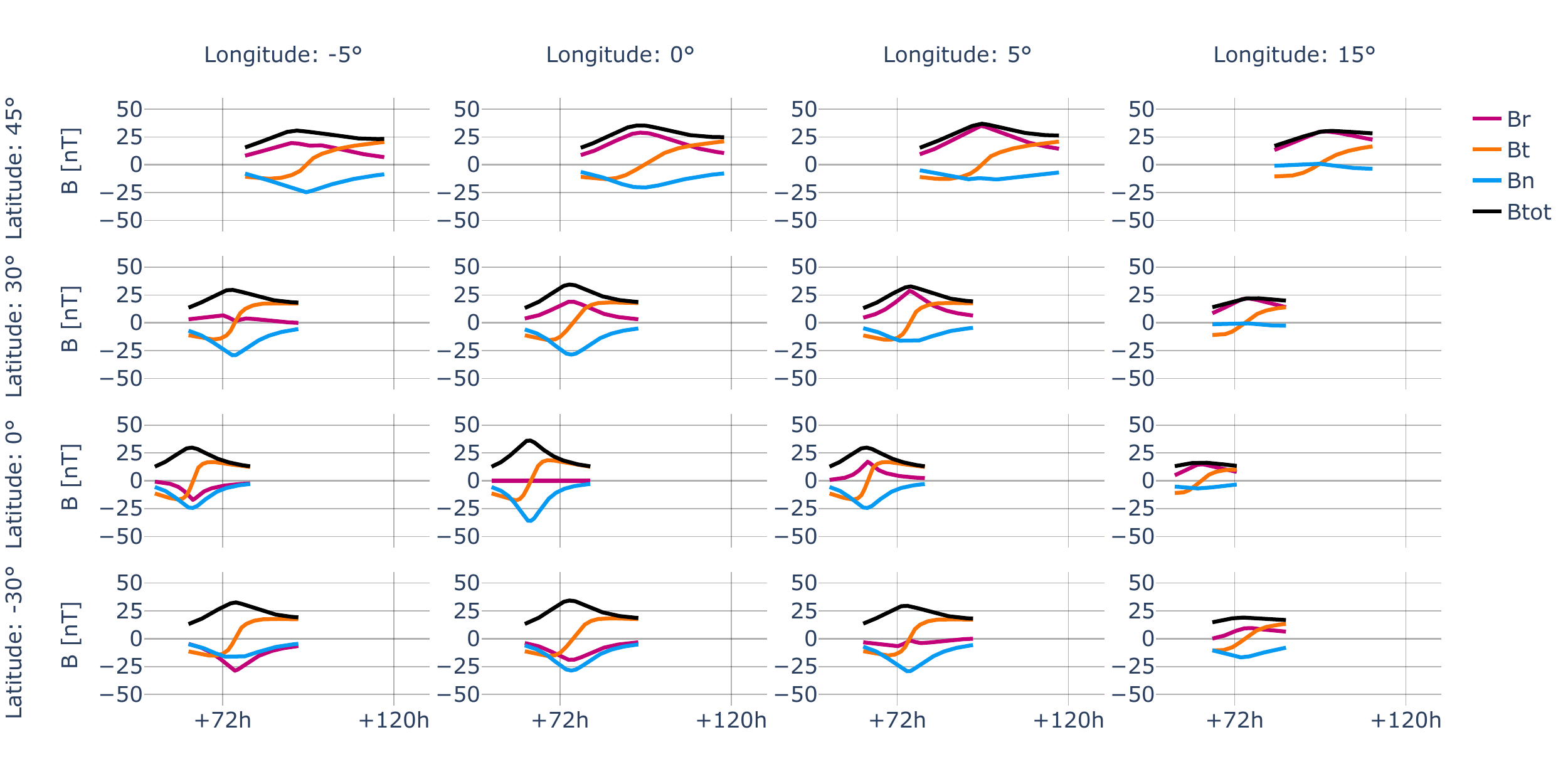}}
\caption{In situ measurements of the synthetic spacecraft in different latitudes and longitudes for a high inclination and high twist ESW type flux rope. For better comparison of the times of arrival, the x-axis displays the same time range for all subplots.} 
\label{fig:high_inc_high_twist_ESW_high}
\end{figure*}

The second right-handed high inclination flux rope type is shown in Figure \ref{fig:high_inc_high_twist_ESW_high}, for which $B_{t}$ changes sign from negative to positive and  $B_{n}$ remains negative.

\newpage

\begin{figure}[h!]
\centering
{\includegraphics[width=0.9\textwidth]{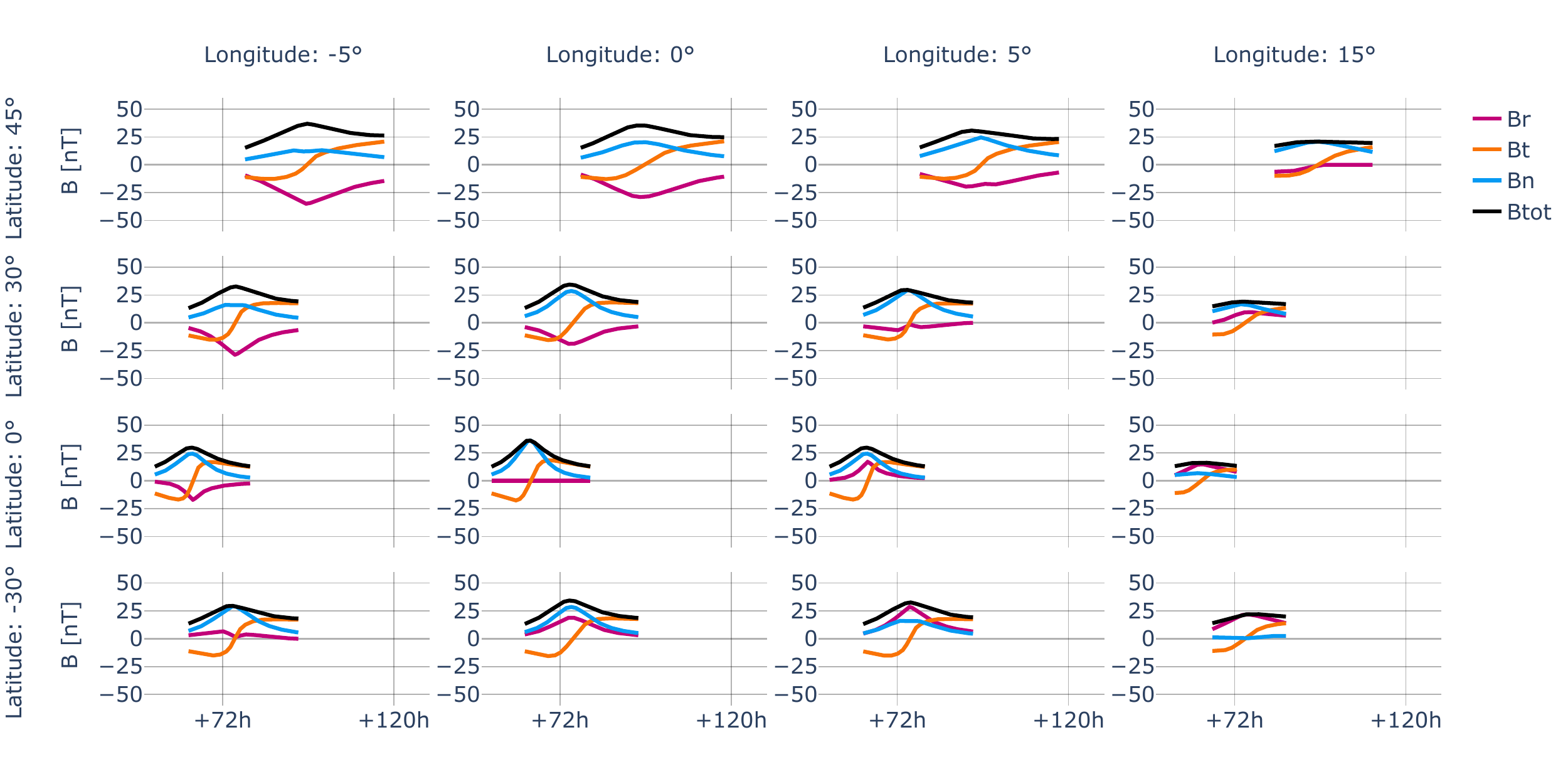}}
\caption{In situ measurements of the synthetic spacecraft in different latitudes and longitudes for a high inclination and high twist ENW type flux rope. For better comparison of the times of arrival, the x-axis displays the same time range for all subplots.} 
\label{fig:high_inc_high_twist_ENW_high}
\end{figure}

\begin{figure}[h!]
\centering
{\includegraphics[width=0.9\textwidth]{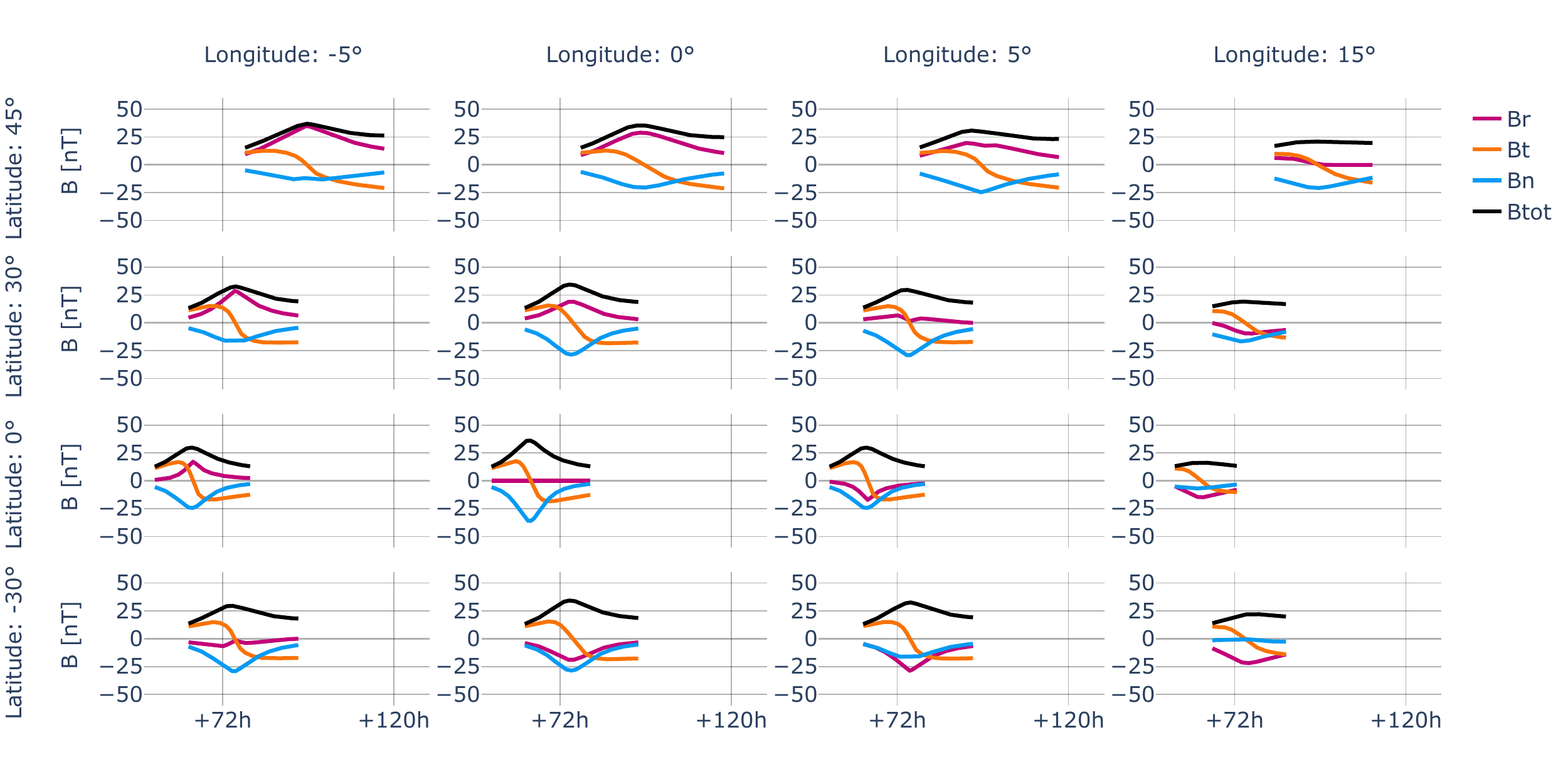}}
\caption{In situ measurements of the synthetic spacecraft in different latitudes and longitudes for a high inclination and high twist WSE type flux rope. For better comparison of the times of arrival, the x-axis displays the same time range for all subplots.} 
\label{fig:high_inc_high_twist_WSE_high}
\end{figure}

Figures \ref{fig:high_inc_high_twist_ENW_high} and \ref{fig:high_inc_high_twist_WSE_high} show the left-handed counterparts to these, where the ENW type exhibits $B_{t}$ changing sign from negative to positive and $B_{n}$ remaining positive, while the WSE type has $B_{t}$ changing sign from positive to negative and $B_{n}$ remaining negative.

\newpage

\begin{figure*}[h!]
\centering
{\includegraphics[width=\textwidth]{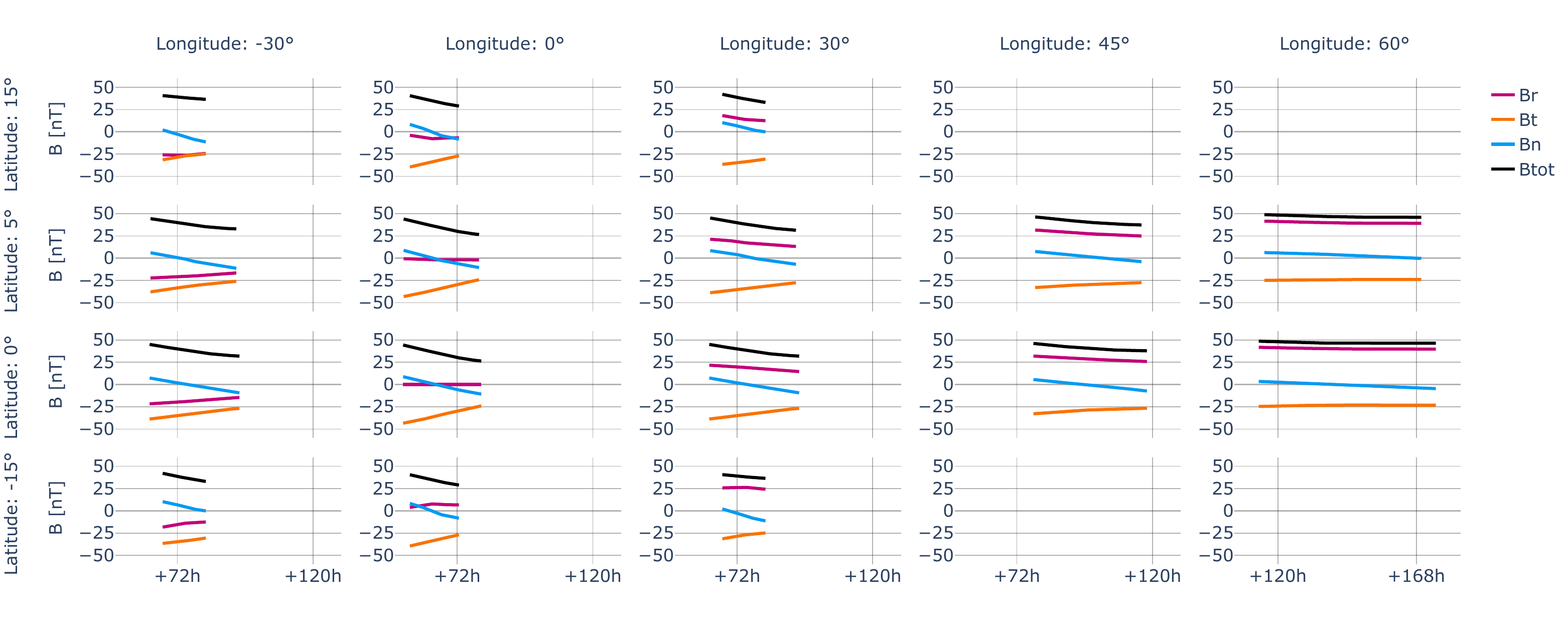}}
\caption{In situ measurements of the synthetic spacecraft in different latitudes and longitudes for a low inclination and low twist NES type flux rope. For better comparison of the times of arrival, the x-axis displays the same time range for all subplots except for a longitude of $60\dgg$.} 
\label{fig:low_inc_low_twist_NES_low}
\end{figure*}

Figure \ref{fig:low_inc_low_twist_NES_low} shows the second right-handed low inclination flux rope type NES $B_{n}$ changes sign from slightly positive to slightly negative and  $B_{t}$ remains negative.

\newpage

\begin{figure*}[h!]
\centering
{\includegraphics[width=0.9\textwidth]{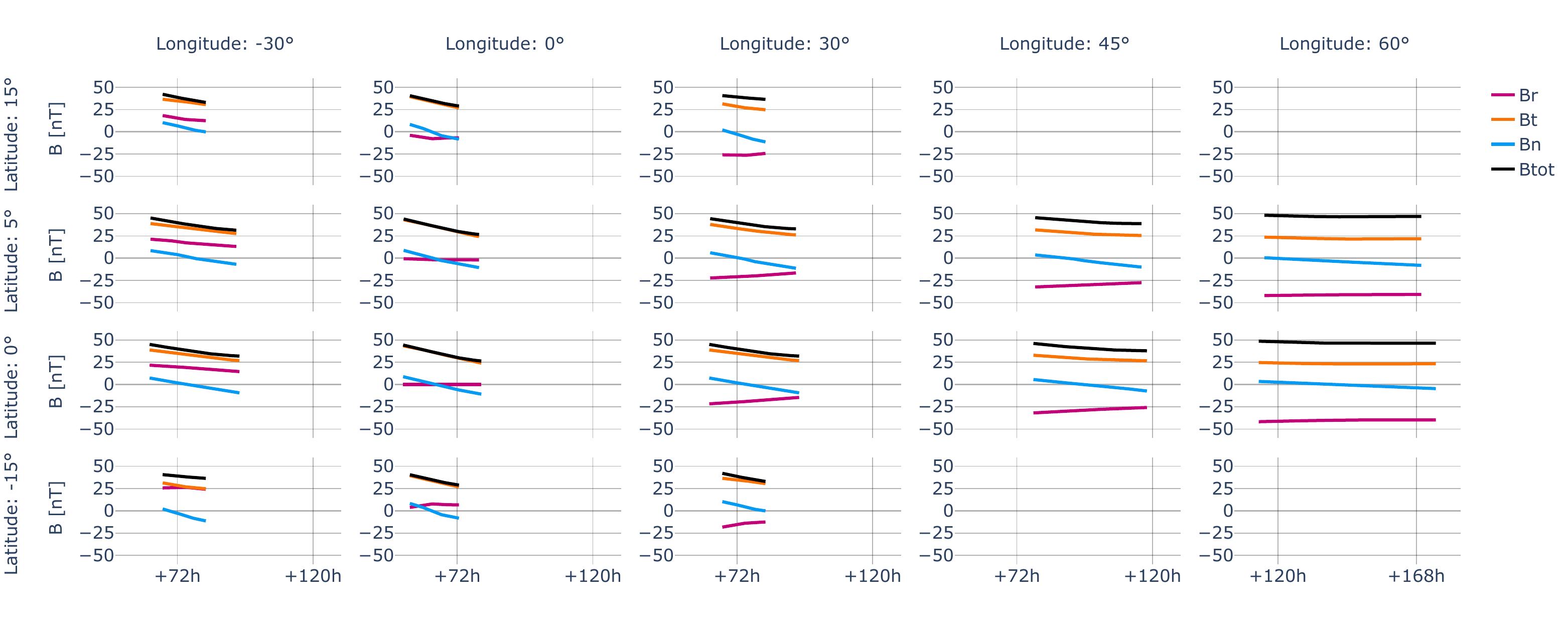}}
\caption{In situ measurements of the synthetic spacecraft in different latitudes and longitudes for a low inclination and low twist NWS type flux rope. For better comparison of the times of arrival, the x-axis displays the same time range for all subplots except for a longitude of $60\dgg$.} 
\label{fig:low_inc_low_twist_NWS_low}
\end{figure*}

\begin{figure*}[h!]
\centering
{\includegraphics[width=0.9\textwidth]{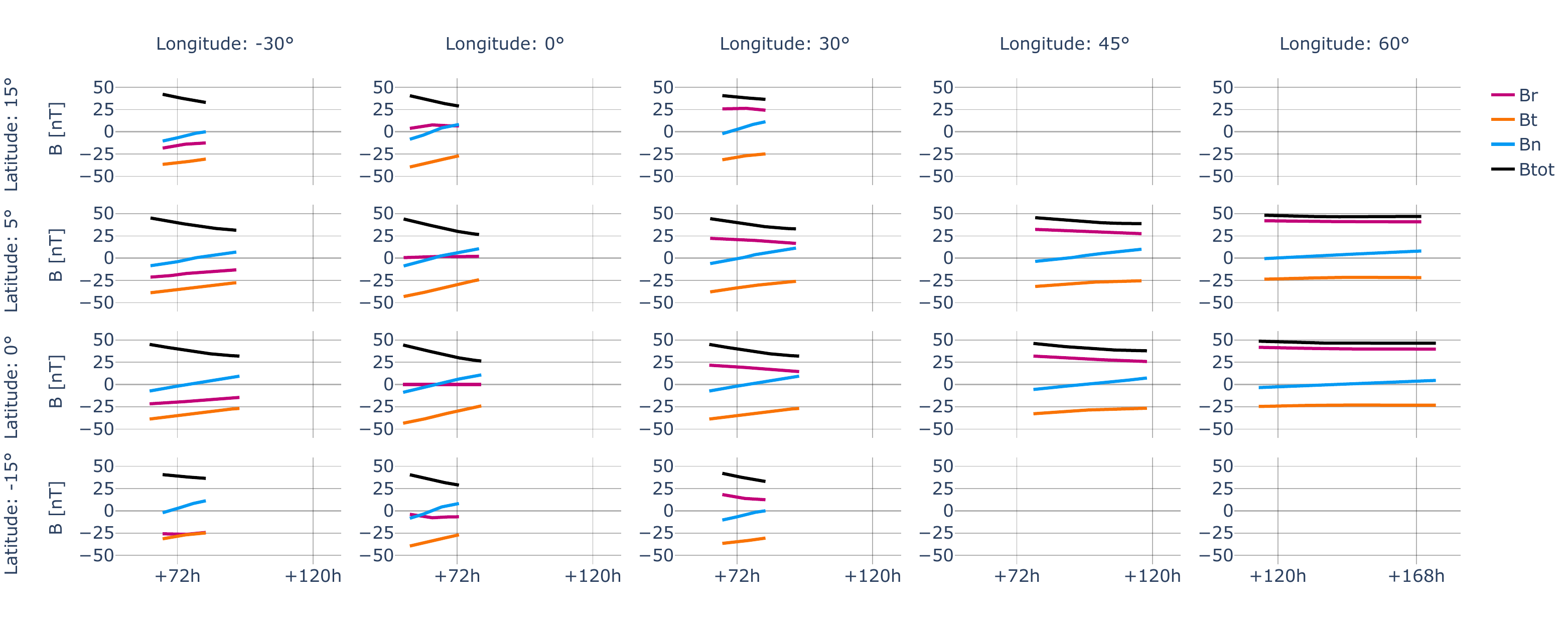}}
\caption{In situ measurements of the synthetic spacecraft in different latitudes and longitudes for a low inclination and low twist SEN type flux rope. For better comparison of the times of arrival, the x-axis displays the same time range for all subplots except for a longitude of $60\dgg$.} 
\label{fig:low_inc_low_twist_SEN_low}
\end{figure*}

Figures \ref{fig:low_inc_low_twist_NWS_low} and \ref{fig:low_inc_low_twist_SEN_low} show the left-handed counterparts to these, where the NWS type exhibits $B_{n}$ changing sign from slightly positive to slightly negative and  $B_{t}$ remaining positive, while the SEN type has $B_{n}$ changing sign from slightly negative to slightly positive and  $B_{t}$ remaining negative.


\newpage

\begin{figure*}[h!]
\centering
{\includegraphics[width=\textwidth]{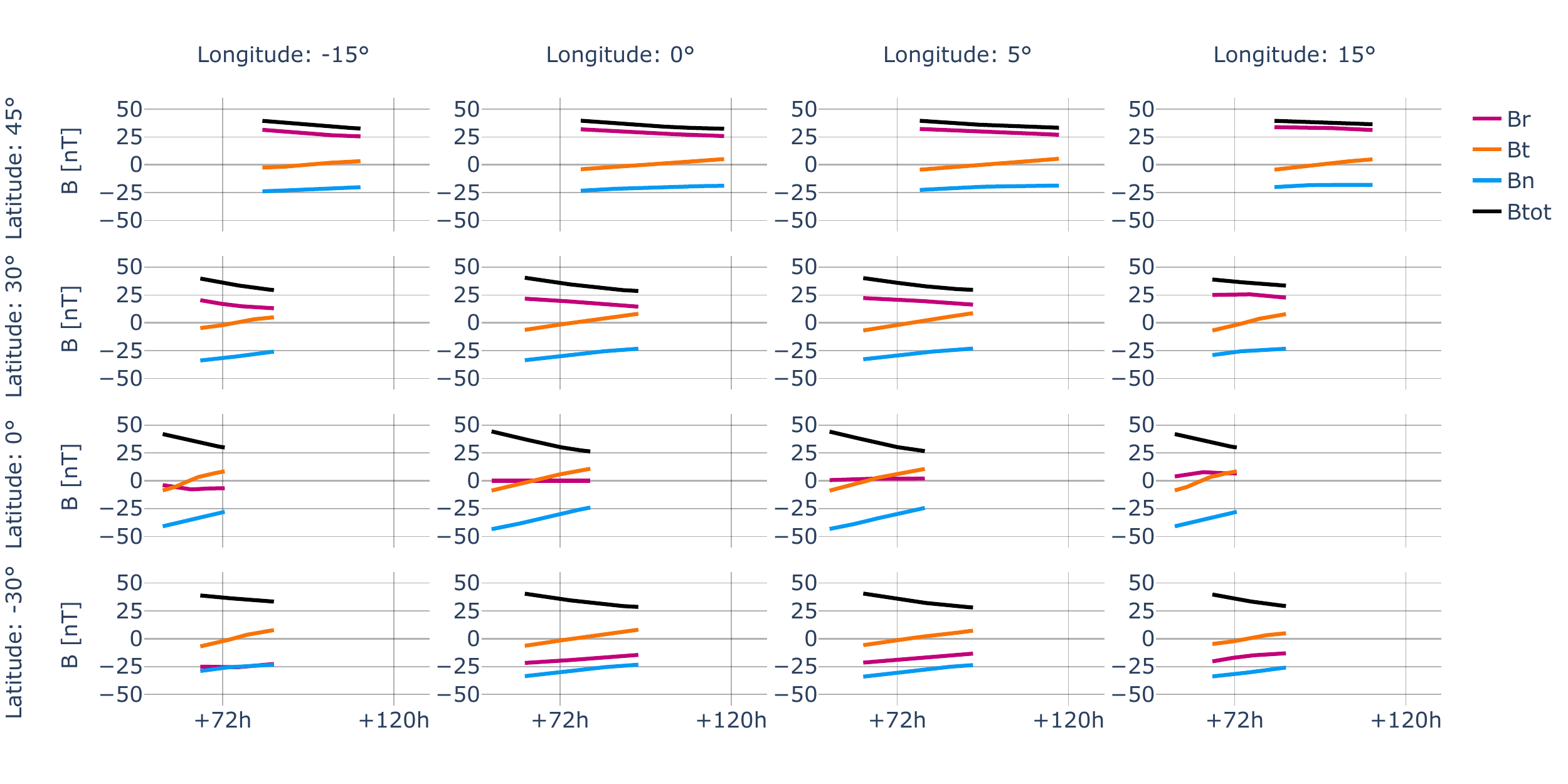}}
\caption{In situ measurements of the synthetic spacecraft in different latitudes and longitudes for a high inclination and low twist ESW type flux rope. For better comparison of the times of arrival, the x-axis displays the same time range for all subplots.} 
\label{fig:high_inc_low_twist_ESW_high}
\end{figure*}

The second right-handed high inclination flux rope type is shown in Figure \ref{fig:high_inc_low_twist_ESW_high}, for which $B_{t}$ changes sign from slightly negative to slightly positive and  $B_{t}$ remains negative.

\newpage

\begin{figure*}[h!]
\centering
{\includegraphics[width=0.9\textwidth]{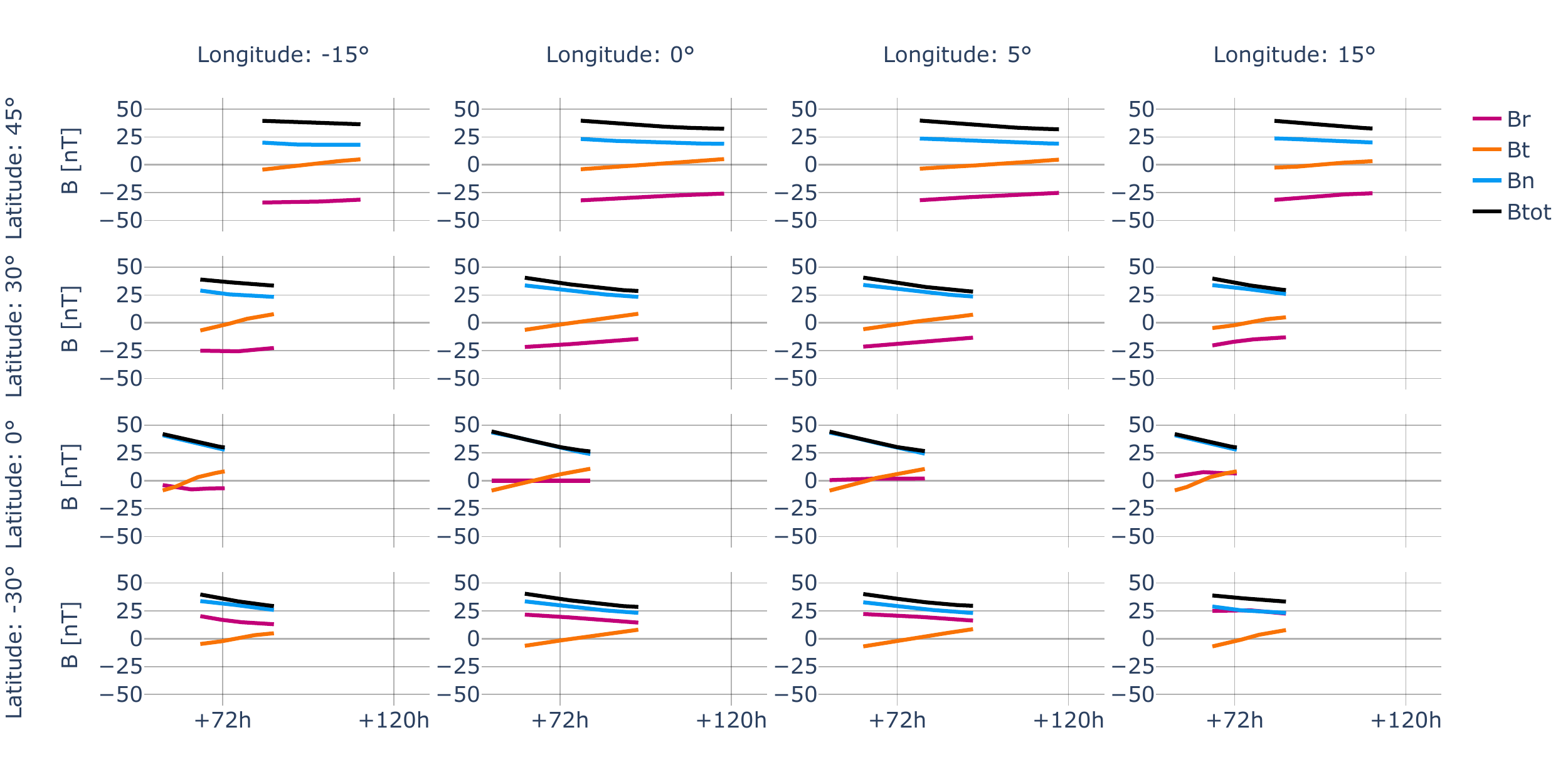}}
\caption{In situ measurements of the synthetic spacecraft in different latitudes and longitudes for a high inclination and low twist ENW type flux rope. For better comparison of the times of arrival, the x-axis displays the same time range for all subplots.} 
\label{fig:high_inc_low_twist_ENW_high}
\end{figure*}

\begin{figure*}[h!]
\centering
{\includegraphics[width=0.9\textwidth]{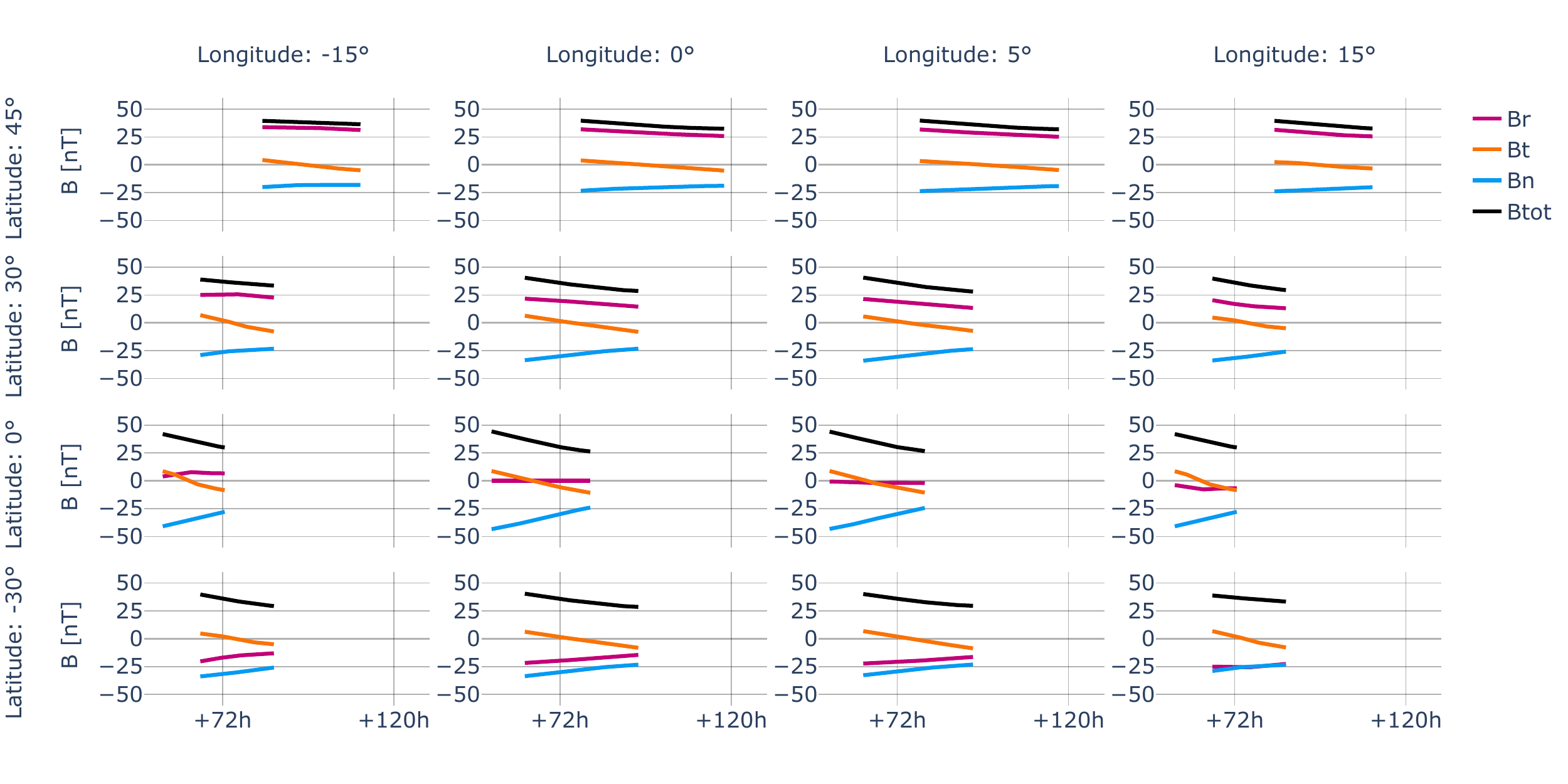}}
\caption{In situ measurements of the synthetic spacecraft in different latitudes and longitudes for a high inclination and low twist WSE type flux rope. For better comparison of the times of arrival, the x-axis displays the same time range for all subplots.} 
\label{fig:high_inc_low_twist_WSE_high}
\end{figure*}

Figures \ref{fig:high_inc_low_twist_ENW_high} and \ref{fig:high_inc_low_twist_WSE_high} show the left-handed counterparts to these, where the ENW type exhibits $B_{t}$ changing sign from slightly negative to slightly positive and $B_{n}$ remaining positive, while the WSE type has $B_{t}$ changing sign from slightly positive to slightly negative and $B_{n}$ remaining negative.

\clearpage


\bibliography{bibliography}{}
\bibliographystyle{aasjournal}

\end{document}